\documentclass[journal]{IEEEtran}
\usepackage{amsmath,amsfonts,amssymb}
\usepackage{array}
\usepackage{graphicx}
\graphicspath{{./}}
\usepackage{cite}
\usepackage{stfloats}
\usepackage{placeins}
\usepackage{balance}
\begin{document}

\title{TwinViT-DeepJSCC: Adversarially Robust Semantic Image Communication}

\author{Maedeh~Fallahreyhani,~\IEEEmembership{Student Member,~IEEE,}
	Paeiz~Azmi,~\IEEEmembership{Senior Member,~IEEE,}
	Nader~Mokari,~\IEEEmembership{Senior Member,~IEEE,}
	M.~Reza~Abedi,~\IEEEmembership{Member,~IEEE,}
	Melike~Erol-Kantarci,~\IEEEmembership{Fellow,~IEEE,}
	and Eduard~A.~Jorswieck,~\IEEEmembership{Fellow,~IEEE}
	\thanks{M. Fallahreyhani, P. Azmi, N. Mokari, and M. R. Abedi are with the Department of Electrical Engineering, Tarbiat Modares University, Tehran, Iran (e-mail: m.fallahreyhani@modares.ac.ir; pazmi@modares.ac.ir; nader.mokari@modares.ac.ir; mohammadreza\_abedi@modares.ac.ir).}
	\thanks{M. Erol-Kantarci is with the School of Electrical Engineering and Computer Science, University of Ottawa, Ottawa, Canada (e-mail: merolka2@uottawa.ca).}
	\thanks{E. A. Jorswieck is with the Institute for Communications Technology, Technische Universit\"at Braunschweig, Braunschweig, Germany (e-mail: e.jorswieck@tu-braunschweig.de).}
}

\markboth{Preprint submitted to IEEE Transactions on Communications}%
{Fallahreyhani \MakeLowercase{\textit{et al.}}: TwinViT-DeepJSCC for Robust Semantic Image Communication}

\maketitle

\begin{center}
\footnotesize\itshape
This work has been submitted to the IEEE for possible publication. Copyright may be transferred without notice, after which this version may no longer be accessible.
\end{center}
\vspace{0.5em}

\begin{abstract}
	Learning-based semantic communication is vulnerable to adversarial perturbations introduced before semantic encoding or over wireless channels. This paper proposes TwinViT-DeepJSCC, a preventive--corrective semantic image transceiver operating under a fixed channel-use budget. Two Vision Transformer (ViT)-based deep joint source--channel coding (DeepJSCC) branches learn complementary latent representations protected by sensitivity-aware masking. At the receiver, confidence-aware fusion, blind corruption-severity estimation, and signal-to-noise ratio (SNR)--severity-conditioned denoising diffusion implicit model (DDIM) purification mitigate residual corruption without requiring attack metadata. Experiments on the Canadian Institute for Advanced Research 100-class (CIFAR-100) dataset consider fast gradient sign method (FGSM), projected gradient descent (PGD), natural evolution strategies (NES), and Carlini--Wagner (CW) source-domain attacks, as well as random jamming and channel-aware adversarial waveforms over additive white Gaussian noise (AWGN) and block-flat Rayleigh fading. Under matched channel-use and attack budgets, TwinViT-DeepJSCC achieves maximum peak signal-to-noise ratio (PSNR) gains of approximately 9.5~dB under 20-step PGD and 10.8~dB under channel-aware waveform attacks over block-flat Rayleigh fading. Under PGD, it also improves Top-1 accuracy by up to approximately 38 percentage points over the undefended baseline and 13 percentage points over the strongest competing defense. Ablation results confirm the complementary contributions of the proposed transmitter- and receiver-side mechanisms.
\end{abstract}

\begin{IEEEkeywords}
	Semantic communications, DeepJSCC, Vision Transformer, adversarial robustness, latent-space protection, diffusion purification.
\end{IEEEkeywords}
\section{Introduction}
\label{sec:introduction}

Semantic communication shifts the communication objective from exact bit recovery toward reliable delivery of task-relevant meaning~\cite{xie2021deeplearning,luo2022semantic,niu2022paradigm,dai2023beyond}. Realizing this objective is challenging with conventional communication architectures because task-relevant semantic features, source statistics, channel conditions, and downstream inference objectives are generally coupled in a highly nonlinear manner. This motivates learning-based transceivers that can jointly optimize semantic representation, channel mapping, and reconstruction according to the end-to-end communication objective. Deep joint source--channel coding (DeepJSCC) realizes this paradigm through end-to-end learning of source representation, compression, channel protection, and reconstruction, thereby providing graceful degradation under bandwidth constraints, channel mismatch, and low signal-to-noise ratio (SNR) conditions~\cite{bourtsoulatze2019deepjscc,kurka2020deepjscc,dai2022nonlinear}. Recent extensions based on attention mechanisms, Transformers, orthogonal frequency-division multiplexing (OFDM), and multiple-input multiple-output (MIMO) transmission have further improved rate--distortion performance and channel adaptation~\cite{xu2022attention,yang2023witt,yang2025swinjscc,wu2024feedback,wu2024mimo}; task-oriented DeepJSCC has also been developed to jointly support image recovery and classification~\cite{lyu2024recoveryclassification}. However, robustness to stochastic channel impairments does not necessarily imply robustness to deliberately optimized adversarial perturbations.

Learning-based semantic transceivers are vulnerable to source-domain perturbations introduced before semantic encoding and channel-domain attacks that manipulate transmitted latent symbols or receiver observations~\cite{sagduyu2023multidomain,nan2023physical,liu2024manipulating}. This work focuses on active integrity attacks that alter the semantic information delivered to the legitimate receiver; confidentiality-, privacy-, model-inversion-, and poisoning-oriented threats are outside its scope. Here, \emph{semantic robustness} denotes preservation of both reconstruction fidelity and downstream task performance under bounded adversarial perturbations, stochastic channel impairments, and an unchanged total channel-use budget. Since high pixel fidelity does not necessarily preserve class-discriminative information, robustness is assessed using both peak signal-to-noise ratio (PSNR) and Top-1 classification accuracy.

Existing defenses protect semantic links through robust latent coding and feature selection~\cite{hu2023maskedvqvae,peng2024robustimage}, adversarial training~\cite{liu2023semprotector,shi2025securedsc}, or receiver-side filtering, ensembling, and diffusion purification~\cite{zhou2026rome,ren2024diffusion,zhao2026secdiff}. However, these methods typically address a specific attack surface or defense location and do not jointly combine transmitter-side latent protection, reliability-aware fusion, blind corruption-severity estimation, and adaptive purification under a fixed channel-use budget. Moreover, inherent robustness to stochastic channel noise should be distinguished from explicit protection against adaptive adversarial attacks~\cite{zhang2026unanticipated}.

Motivated by these limitations, we propose \emph{TwinViT-DeepJSCC}, a preventive--corrective framework for adversarially robust semantic image communication. Under a fixed aggregate channel-use budget, two Vision Transformer (ViT)-based DeepJSCC branches learn aligned yet redundancy-reduced latent representations, while sample-adaptive masking suppresses adversarially sensitive components before transmission without requiring attack information at inference. At the receiver, confidence-aware fusion combines the equalized branch observations, and a calibrated blind estimator predicts corruption severity from receiver-observable inter-branch discrepancy, fusion uncertainty, and channel SNR. The estimated severity and SNR then control the starting depth of conditional denoising diffusion implicit model (DDIM) purification, enabling stronger correction for severe corruption while avoiding unnecessarily deep denoising under benign or mild conditions.

The main contributions of this work are summarized as follows:
\begin{itemize}
	\item We formulate a unified source--channel adversarial setting for semantic image communication and develop a preventive--corrective transceiver that protects both transmitted latent representations and reconstructed semantic content under a fixed total channel-use budget.
	
	\item We introduce twin ViT-based DeepJSCC branches with semantic-alignment and redundancy-reduction objectives, together with sample-adaptive masking that suppresses adversarially sensitive latent components without requiring attack information or gradients at inference.
	
	\item We design an attack-agnostic receiver comprising confidence-aware fusion, calibrated blind corruption-severity estimation, and SNR--severity-conditioned DDIM purification with an adaptive denoising depth.
	
	\item We evaluate reconstruction and task-level robustness on the Canadian Institute for Advanced Research 100-class (CIFAR-100) dataset under fast gradient sign method (FGSM), projected gradient descent (PGD), natural evolution strategies (NES), and Carlini--Wagner (CW) source-domain attacks, random and channel-aware interference, additive white Gaussian noise (AWGN), block-flat Rayleigh fading, and multiple perturbation budgets. Module-level and architectural ablations quantify the contributions of the proposed components.
\end{itemize}

The remainder of this paper is organized as follows. Section~II reviews related work, Section~III presents the system and threat models, Section~IV details TwinViT-DeepJSCC, Section~V reports the numerical results, and Section~VI concludes the paper.
\section{Related Works}
\label{sec:related_works}

\subsection{DeepJSCC and Transformer-Based Semantic Image Transmission}
\label{subsec:rw_deepjscc}

DeepJSCC directly maps source images to channel symbols through an end-to-end learned encoder--decoder and provides graceful reconstruction degradation under bandwidth limitation and low-SNR conditions~\cite{bourtsoulatze2019deepjscc}. A VAE-based joint coding--modulation framework has also been developed for digital image semantic communication, jointly adapting learned discrete signaling to channel conditions~\cite{bo2024jcm}. Transformer-based extensions such as the wireless image transmission transformer (WITT) and Swin Transformer-based DeepJSCC (SwinJSCC) improve content and channel adaptation by exploiting long-range spatial dependencies~\cite{yang2023witt,yang2025swinjscc}. The DeepSC-RI system is particularly relevant because it combines dual-branch feature extraction, semantic-importance evaluation, and cross-attention fusion for robust image transmission~\cite{peng2024robustimage}. In contrast, TwinViT-DeepJSCC learns aligned yet redundancy-reduced ViT representations under a fixed aggregate channel-use budget and integrates them with sensitivity-aware masking, reliability-aware fusion, blind severity estimation, and adaptive diffusion purification.

\subsection{Attack Surfaces in Semantic Communications}
\label{subsec:rw_attacks}

Security analyses distinguish attacks on semantic communication systems according to whether perturbations target the source input, learned representation, wireless channel, or shared semantic knowledge~\cite{sagduyu2023multidomain,shen2024secure}. Source-domain attacks construct bounded perturbations using gradient-, optimization-, or query-based methods~\cite{goodfellow2015fgsm,madry2018pgd,carlini2017cw,ilyas2018blackbox}, whereas channel-domain adversaries inject structured interference or optimized waveforms into the wireless link~\cite{nan2023physical,li2022sembat,liu2024manipulating}. This work focuses on active source- and channel-domain integrity attacks that degrade reconstruction and task-level semantic performance; privacy, eavesdropping, and knowledge-manipulation attacks pursue different security objectives and are outside its scope.
\subsection{Robust and Secure Defense Mechanisms}
\label{subsec:rw_defenses}

Existing defenses can be broadly divided into transmitter-side protection and receiver-side correction. Masked vector-quantized variational autoencoder (VQ-VAE) and SemProtector protect semantic representations through selective coding or robust training, while semantic-importance-aware feature allocation and adversarial training further improve the protection of informative representations~\cite{hu2023maskedvqvae,liu2023semprotector,zhou2025featureallocation,shi2025securedsc}. Receiver-side methods instead perform attack detection, adaptive filtering, expert selection, classifier ensembling, or online defense adaptation; representative examples include Secure Semantic Communications via Robust Adversarial Knowledge Learning (SECO), Mixture of Experts-Based Semantic Communication (MESC), Robust Model Ensembling (ROME), and the practical ``plugging and breathing'' framework~\cite{hoang2025seco,he2026mesc,zhou2026rome,qiu2025plugging}. Although these approaches improve robustness within their targeted settings, transmitter-only methods may remain sensitive to attack mismatch, whereas receiver-only defenses act after vulnerable latent features have already been transmitted. TwinViT-DeepJSCC therefore combines preventive latent protection with adaptive receiver-side correction under a unified source--channel threat model.

\subsection{Diffusion-Based Recovery and Adversarial Purification}
\label{subsec:rw_diffusion}

Diffusion models have been applied to generative reconstruction, inverse problems, and adversarial purification~\cite{ho2020ddpm,song2021ddim,nie2022diffpure}. In semantic communication, channel denoising diffusion models (CDDM) and the DiffCom framework condition diffusion-based recovery on channel observations, while diffusion-driven semantic communication further exploits generative priors under bandwidth constraints~\cite{wu2024cddm,wang2025diffcom,guo2025diffusiondriven}. Security-oriented methods are more closely related to this work: receiver-side diffusion purification mitigates source-domain perturbations, and SecDiff addresses pilot spoofing and subcarrier jamming through diffusion-guided inpainting and channel refinement~\cite{ren2024diffusion,zhao2026secdiff}. Unlike these methods, TwinViT-DeepJSCC jointly combines transmitter-side latent masking, confidence-aware fusion, blind severity estimation, and SNR--severity-conditioned DDIM purification for both source- and channel-domain attacks under a fixed channel-use budget.

Table~\ref{tab:related_comparison} positions TwinViT-DeepJSCC against representative robust semantic communication schemes in terms of attack coverage, defense location, online adaptation, and preventive--corrective integration. Unlike the compared methods, the proposed framework jointly integrates transmitter-side latent protection with blind, reliability-aware, and severity-adaptive receiver correction under a fixed channel-use budget.
\begin{table*}[!t]
	\centering
	\caption{Comparison of Representative Adversarially Robust and Secure Semantic Communication Studies}
	\label{tab:related_comparison}
	\scriptsize
	\setlength{\tabcolsep}{2.7pt}
	\renewcommand{\arraystretch}{1.18}
	\resizebox{\textwidth}{!}{%
		\begin{tabular}{p{3.35cm}ccccccccc}
			\hline
			\textbf{Work} &
			\shortstack{\textbf{Image}\\\textbf{DeepJSCC}} &
			\shortstack{\textbf{Source}\\\textbf{Attack}} &
			\shortstack{\textbf{Channel}\\\textbf{Attack}} &
			\shortstack{\textbf{Transmitter-Side}\\\textbf{Prevention}} &
			\shortstack{\textbf{Receiver-Side}\\\textbf{Correction}} &
			\shortstack{\textbf{Blind/Online}\\\textbf{Adaptation}} &
			\shortstack{\textbf{Diffusion}\\\textbf{Defense}} &
			\shortstack{\textbf{Source--Channel}\\\textbf{Joint Coverage}} &
			\shortstack{\textbf{Preventive--}\\\textbf{Corrective}} \\
			\hline
			
			Multi-domain attack analysis~\cite{sagduyu2023multidomain}
			& $\checkmark$ & $\checkmark$ & $\checkmark$ & -- & -- & -- & -- & $\checkmark$ & -- \\
			
			Physical-layer adversarial robustness~\cite{nan2023physical}
			& $\checkmark$ & -- & $\checkmark$ & -- & $\checkmark$ & -- & -- & -- & -- \\
			
			DeepSC-RI~\cite{peng2024robustimage}
			& $\checkmark$ & $\checkmark$ & -- & $\checkmark$ & -- & -- & -- & -- & -- \\
			
			Masked VQ-VAE~\cite{hu2023maskedvqvae}
			& $\checkmark$ & $\checkmark$ & -- & $\checkmark$ & -- & -- & -- & -- & -- \\
			
			SemProtector~\cite{liu2023semprotector}
			& $\checkmark$ & $\checkmark$ & -- & $\checkmark$ & -- & -- & -- & -- & -- \\
			
			SECO~\cite{hoang2025seco}
			& -- & $\checkmark$ & $\checkmark$ & -- & $\checkmark$ & $\triangle$ & -- & $\checkmark$ & -- \\
			
			MESC~\cite{he2026mesc}
			& $\triangle$ & $\checkmark$ & $\checkmark$ & $\triangle$ & $\triangle$ & $\checkmark$ & -- & $\checkmark$ & $\triangle$ \\
			
			ROME~\cite{zhou2026rome}
			& $\triangle$ & -- & $\checkmark$ & -- & $\checkmark$ & $\checkmark$ & -- & -- & -- \\
			
			DiffuSeC~\cite{ren2023diffusec}
			& $\checkmark$ & $\checkmark$ & $\triangle$ & -- & $\checkmark$ & $\checkmark$ & $\checkmark$ & $\triangle$ & -- \\
			
			Diffusion purification~\cite{ren2024diffusion}
			& $\checkmark$ & $\checkmark$ & -- & -- & $\checkmark$ & -- & $\checkmark$ & -- & -- \\
			
			SecDiff~\cite{zhao2026secdiff}
			& $\checkmark$ & -- & $\checkmark$ & -- & $\checkmark$ & $\checkmark$ & $\checkmark$ & -- & -- \\
			
			\textbf{Proposed TwinViT-DeepJSCC}
			& $\checkmark$ & $\checkmark$ & $\checkmark$ & $\checkmark$ & $\checkmark$
			& $\checkmark$ & $\checkmark$ & $\checkmark$ & $\checkmark$ \\
			\hline
	\end{tabular}}
	
	\vspace{1mm}
	\parbox{0.98\textwidth}{\scriptsize
		$\checkmark$: explicitly supported; $\triangle$: partially supported, modular, task-dependent, or not the primary design objective; --: not considered. ``Blind/online adaptation'' denotes inference-time adaptation without prior knowledge of the exact attack type and perturbation budget. The classifications reflect the primary scope and mechanisms reported by the corresponding studies.}
\end{table*}

\section{System Model}
\label{sec:system_model}

\begin{figure}[!t]
\centering
\includegraphics[width=0.98\columnwidth]{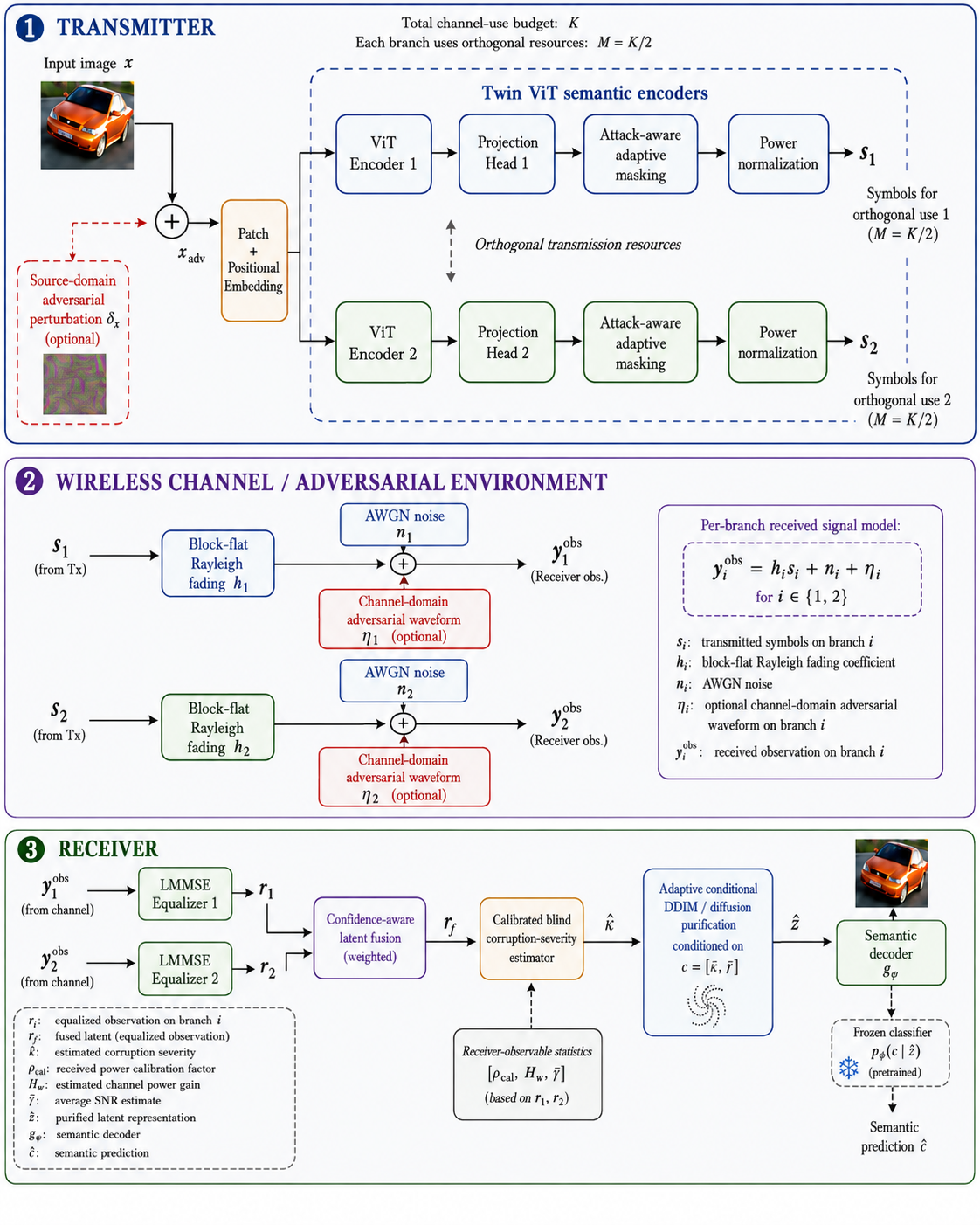}
\caption{End-to-end architecture of TwinViT-DeepJSCC. Two ViT-based semantic branches generate aligned yet redundancy-reduced latent representations. Sample-adaptive masking is applied before transmission over orthogonal resources under a fixed total channel-use budget. The receiver performs legitimate-channel equalization, confidence-aware fusion, calibrated blind corruption-severity estimation, adaptive conditional DDIM purification, semantic reconstruction, and downstream inference.}
	\label{fig:system_model}
\end{figure}

This section formalizes the point-to-point semantic image communication system considered in this work. A source image may be adversarially perturbed before semantic encoding, while the transmitted latent symbols may subsequently be affected by stochastic channel impairments or a deliberately designed channel-domain waveform. Throughout the paper, the total number of complex channel uses is held fixed across all competing schemes. The complete end-to-end architecture is illustrated in Fig.~\ref{fig:system_model}.

\begin{table*}[!t]
	\centering
	\caption{Main Symbols Used in the Paper}
	\label{tab:symbols}
	\scriptsize
	\setlength{\tabcolsep}{3.6pt}
	\renewcommand{\arraystretch}{1.13}
	\begin{tabular}{p{0.09\textwidth}p{0.38\textwidth}p{0.09\textwidth}p{0.38\textwidth}}
		\hline
		\textbf{Symbol} & \textbf{Definition} &
		\textbf{Symbol} & \textbf{Definition} \\
		\hline
		
		$\mathbf{x},\mathbf{x}_{\mathrm{adv}}$
		& Clean and source-domain adversarial images. &
		$H,W,C$
		& Image height, width, and number of channels. \\
		
		$K,M$
		& Total and per-branch complex channel uses, with $M=K/2$. &
		$\mathcal{E}_{\theta_i},\mathcal{P}_{\omega_i}$
		& ViT encoder and projection head of branch $i$. \\
		
		$\mathbf{z}_i,\mathbf{v}_i$
		& Complex latent vector and its real-valued representation. &
		$\mathbf{s}_i,P_t$
		& Transmitted vector and nominal average power per channel use. \\
		
		$h_i,\widehat h_i$
		& Legitimate channel coefficient and its receiver-side estimate. &
		$\mathbf{n}_i,\sigma_n^2$
		& Complex Gaussian noise vector and per-component variance. \\
		
		$\mathbf{y}_i,\mathbf{y}_i^{\mathrm{adv}}$
		& Benign and adversarially corrupted received vectors. &
		$\mathbf{y}_i^{\mathrm{obs}}$
		& Receiver observation under either benign or adversarial operation. \\
		
		$\boldsymbol{\delta}_x,\epsilon_x$
		& Source perturbation and its norm budget. &
		$\boldsymbol{\eta}_i,g_i$
		& Channel-domain adversarial waveform and adversarial channel coefficient. \\
		
		$P_{\mathrm{adv}},\mathrm{ASR}$
		& Adversarial transmit power and nominal adversarial-to-signal power ratio. &
		$\mathbf{r}_i,\mathbf{r}_f$
		& Equalized branch observation and fused latent representation. \\
		
		$\mathbf{C}_{12},\mathcal{L}_{\mathrm{twin}}$
		& Cross-correlation matrix and twin-representation loss. &
		$\boldsymbol{\pi}_i,\mathbf{m}_i$
		& Retention-probability vector and binary latent mask. \\
		
		$\chi_{i,k},q$
		& Sensitivity of component $k$ and target retention ratio. &
		$\mathcal{L}_{\mathrm{mask}}$
		& Sensitivity-aware masking objective. \\
		
		$\ell_i,w_i$
		& Branch reliability logit and normalized fusion weight. &
		$\rho,\rho_{\mathrm{cal}}$
		& Raw and SNR-calibrated inter-branch discrepancy. \\
		
		$\mathcal{H}_w$
		& Normalized entropy of the fusion weights. &
		$\kappa,\widehat\kappa$
		& Target and estimated corruption severity. \\
		
		$\bar{\gamma}_i,\bar{\gamma}$
		& Normalized branch SNR and normalized average SNR. &
		$t_{\mathrm{start}}$
		& Adaptive starting depth of DDIM purification. \\
		
		$t_{\min},t_{\max}$
		& Minimum and maximum admissible diffusion depths. &
		$\lambda_\kappa,\lambda_\gamma$
		& Severity and SNR weights in the depth scheduler. \\
		
		$\alpha_t,\bar{\alpha}_t$
		& Diffusion coefficient and its cumulative product. &
		$\boldsymbol{\epsilon}_{\phi}(\cdot)$
		& Conditional diffusion noise-prediction network. \\
		
		$\mathbf{c}$
		& Diffusion condition $[\widehat\kappa,\bar{\gamma}]^T$. &
		$\widehat{\mathbf{z}},\widehat{\mathbf{x}}$
		& Purified latent vector and reconstructed image. \\
		
		$\mathcal{L}_{\mathrm{rec}},\mathcal{L}_{\mathrm{sem}}$
		& Reconstruction and semantic-task losses. &
		$\mathcal{L}_{\mathrm{diff}},\mathcal{L}_{\mathrm{total}}$
		& Diffusion loss and complete training objective. \\
		
		\hline
	\end{tabular}
\end{table*}

\subsection{Semantic Transceiver and Channel-Use Budget}
\label{subsec:semantic_transceiver}

Let
$\mathbf{x}\in[0,1]^{H\times W\times C}$
denote a source image drawn from the distribution
$p_{\mathcal{X}}$. The complete transmission is constrained to
$K$ complex channel uses, where $K$ is assumed to be even. The two branches are assigned equal and mutually orthogonal communication resources, such that
\begin{equation}
	M=\frac{K}{2}
\end{equation}
complex channel uses are allocated to each branch. Consequently, the overall channel-bandwidth ratio is
$K/(HWC)$, which is identical to that of the single-branch baselines used for comparison.

For branch $i\in\{1,2\}$, let $\mathcal E_{\theta_i}$ and $\mathcal P_{\omega_i}$ denote the ViT encoder and projection head parameterized by $\theta_i$ and $\omega_i$, respectively. They generate
\begin{equation}
	\mathbf{z}_i
	=
	\mathcal{P}_{\omega_i}
	\left(
	\mathcal{E}_{\theta_i}(\mathbf{x})
	\right)
	\in\mathbb{C}^{M},
	\qquad i\in\{1,2\}.
	\label{eq:branch_latent}
\end{equation}
The projection head includes the deterministic pairing of
$2M$ real-valued outputs into $M$ complex latent symbols. Both projection heads map their corresponding branch features into a common-dimensional coordinate space, enabling subsequent branch-wise comparison and fusion.

After sample-adaptive masking and power normalization, branch $i$ transmits
$\mathbf{s}_i\in\mathbb{C}^{M}$ subject to
\begin{equation}
	\frac{1}{M}
	\mathbb{E}
	\left[
	\|\mathbf{s}_i\|_2^2
	\right]
	\leq P_t,
	\qquad i\in\{1,2\},
	\label{eq:transmit_power_constraint}
\end{equation}
where $P_t$ is the nominal average transmit-power budget per complex channel use and the expectation is taken over source samples and any stochastic operations used during training. Since both branches use $M=K/2$ symbols, the twin architecture does not obtain additional bandwidth or channel uses relative to the single-branch baselines.

\subsection{Wireless Channel Model}
\label{subsec:wireless_channel}

The two latent vectors are transmitted over orthogonal time, frequency, or code resources. Under benign operation, the received vector associated with branch $i$ is
\begin{equation}
	\mathbf{y}_i
	=
	h_i\mathbf{s}_i+\mathbf{n}_i,
	\qquad i\in\{1,2\},
	\label{eq:benign_channel}
\end{equation}
where $h_i\in\mathbb{C}$ denotes the legitimate scalar channel coefficient and
\begin{equation}
	\mathbf{n}_i
	\sim
	\mathcal{CN}
	\left(
	\mathbf{0},
	\sigma_n^2\mathbf{I}_M
	\right).
	\label{eq:channel_noise}
\end{equation}
Here, $\mathcal{CN}$ denotes a proper complex Gaussian distribution, $\sigma_n^2$ is the per-component noise variance, and $\mathbf I_M$ is the $M\times M$ identity matrix. The noise vectors of the two orthogonal branches are assumed to be statistically independent and independent of the source, channel coefficients, and adversarial waveforms.

For the additive white Gaussian noise (AWGN) channel,
$h_i=1$. For block-flat Rayleigh fading,
$h_i\sim\mathcal{CN}(0,1)$ and remains constant throughout the transmission of the corresponding $M$-dimensional latent vector. The instantaneous received SNR of branch $i$ is
\begin{equation}
	\gamma_i
	=
	\frac{|h_i|^2P_t}{\sigma_n^2}.
	\label{eq:instantaneous_snr}
\end{equation}
Because $\mathbb{E}[|h_i|^2]=1$ under the adopted block-flat Rayleigh model, the nominal average SNR used on the horizontal axes of the simulation figures is
\begin{equation}
	\gamma
	=
	\frac{P_t}{\sigma_n^2}.
	\label{eq:nominal_snr}
\end{equation}

\subsection{Adversarial Threat Model}
\label{subsec:adversarial_threats}

We consider active integrity attacks introduced at two distinct locations: before semantic encoding and during wireless transmission. The receiver is not provided with the attack family, perturbation budget, adversarial channel parameters, or identity of the attacked branch at inference time.

The considered adversary does not modify the trained transceiver parameters, receiver architecture, or legitimate channel-state estimator. Confidentiality- and privacy-oriented threats, including eavesdropping, model inversion, and semantic leakage, are outside the present scope. Knowledge-base poisoning and training-data poisoning are also excluded. These assumptions isolate attacks that directly corrupt the semantic information delivered through the legitimate communication link.

\subsubsection{Source-Domain Adversarial Perturbations}

A source-domain adversary constructs
\begin{subequations}
	\label{eq:source_domain_attack}
	\begin{align}
		\mathbf{x}_{\mathrm{adv}}
		&=
		\Pi_{[0,1]}
		\left(
		\mathbf{x}
		+
		\boldsymbol{\delta}_x
		\right),
		\label{eq:adv_input}
		\\
		\|\boldsymbol{\delta}_x\|_p
		&\leq
		\epsilon_x,
		\qquad
		p\in\{2,\infty\},
		\label{eq:adv_input_constraint}
	\end{align}
\end{subequations}
where
$\boldsymbol{\delta}_x\in\mathbb{R}^{H\times W\times C}$
is the adversarial perturbation, $\epsilon_x$ is its norm budget, $p$ specifies the $\ell_p$-norm order, and
$\Pi_{[0,1]}(\cdot)$ denotes element-wise projection onto the valid image range. Depending on the attack, the perturbation is optimized to increase the reconstruction loss, the downstream semantic-task loss, or their weighted combination. The same perturbed image is subsequently processed by both semantic branches.

\subsubsection{Channel-Domain Adversarial Waveforms}

A malicious transmitter may inject an adversarial waveform
$\boldsymbol{\eta}_i\in\mathbb{C}^{M}$
through an adversarial channel coefficient
$g_i\in\mathbb{C}$. The resulting received vector is
\begin{subequations}
	\label{eq:channel_attack}
	\begin{align}
		\mathbf{y}_i^{\mathrm{adv}}
		&=
		h_i\mathbf{s}_i
		+
		g_i\boldsymbol{\eta}_i
		+
		\mathbf{n}_i,
		\qquad i\in\{1,2\},
		\label{eq:channel_received}
		\\
		\frac{1}{M}
		\mathbb{E}
		\left[
		\|\boldsymbol{\eta}_i\|_2^2
		\right]
		&\leq
		P_{\mathrm{adv}}.
		\label{eq:channel_attack_constraint}
	\end{align}
\end{subequations}
Here, $P_{\mathrm{adv}}$ is the attacker's average transmit-power budget per channel use. If only one branch is attacked, the waveform associated with the unattacked branch is set to zero. For random Gaussian jamming, the waveform is sampled independently of the legitimate semantic signal. In contrast, a channel-aware adversarial waveform is optimized to increase the reconstruction or semantic-task loss subject to
\eqref{eq:channel_attack_constraint}.

The nominal transmit-side adversarial-to-signal power ratio (ASR) is defined as
\begin{equation}
	\mathrm{ASR}
	=
	10\log_{10}
	\left(
	\frac{P_{\mathrm{adv}}}{P_t}
	\right)
	\ \mathrm{dB}.
	\label{eq:asr}
\end{equation}
This definition deliberately separates the attacker transmit-power budget from the instantaneous channel realizations. For given $h_i$ and $g_i$, the corresponding instantaneous received adversarial-to-signal ratio is
\begin{equation}
	\mathrm{ASR}_{i}^{\mathrm{rx}}
	=
	10\log_{10}
	\left(
	\frac{|g_i|^2P_{\mathrm{adv}}}
	{|h_i|^2P_t}
	\right)
	\ \mathrm{dB}.
	\label{eq:received_asr}
\end{equation}

\subsection{Receiver-Side Channel Equalization}
\label{subsec:channel_equalization}

Let
$\mathbf{y}_i^{\mathrm{obs}}$
denote either the benign observation
$\mathbf{y}_i$
or the adversarially corrupted observation
$\mathbf{y}_i^{\mathrm{adv}}$. The receiver possesses an estimate
$\widehat h_i$
of the legitimate channel coefficient but has no knowledge of
$g_i$, $\boldsymbol{\eta}_i$, or $P_{\mathrm{adv}}$.

Under the nominal benign second-order model
$\mathbb{E}[\mathbf{s}_i\mathbf{s}_i^{H}]
=P_t\mathbf{I}_M$,
the receiver applies the scalar linear minimum mean-square error (LMMSE) coefficient
\begin{equation}
	\mathbf{r}_i
	=
	\frac{
		P_t\widehat h_i^{*}
	}{
		P_t|\widehat h_i|^2+\sigma_n^2
	}
	\mathbf{y}_i^{\mathrm{obs}},
	\qquad i\in\{1,2\}.
	\label{eq:lmmse_equalization}
\end{equation}
Here, $\mathbf r_i$ is the equalized branch observation, $(\cdot)^*$ denotes complex conjugation, and $(\cdot)^H$ denotes Hermitian transpose. The equalizer is termed \emph{nominal} because the adversarial waveform is treated as an unmodeled disturbance and is not incorporated into the equalizer coefficient. Unless otherwise stated, the numerical experiments assume perfect legitimate-channel estimation, i.e.,
$\widehat h_i=h_i$.
\section{The Proposed Method}
\label{sec:proposed_method}

TwinViT-DeepJSCC adopts a preventive--corrective design. At the transmitter, twin representation learning and sample-adaptive latent masking reduce the exposure of vulnerable semantic components before transmission. At the receiver, confidence-aware fusion, calibrated blind corruption-severity estimation, and conditional diffusion purification mitigate residual source- and channel-domain corruption.

\subsection{Aligned and Redundancy-Reduced Twin Representations}
\label{subsec:twin_representation}

Let
\begin{equation}
	\mathbf v_i
	=\mathcal R(\mathbf z_i)
	=
	\big[
	\operatorname{Re}(\mathbf z_i)^T,
	\operatorname{Im}(\mathbf z_i)^T
	\big]^T
	\in\mathbb R^{2M},
	\qquad i\in\{1,2\},
	\label{eq:real_representation}
\end{equation}
where $\mathcal R:\mathbb C^M\rightarrow\mathbb R^{2M}$ denotes the real-valued embedding of a complex latent vector. For a mini-batch of size $B$, let
$\widetilde{\mathbf V}_i\in\mathbb R^{B\times 2M}$
collect the dimension-wise centered and variance-normalized branch representations. The cross-correlation matrix is
\begin{equation}
	\mathbf C_{12}
	=
	\frac{1}{B}
	\widetilde{\mathbf V}_1^T
	\widetilde{\mathbf V}_2
	\in\mathbb R^{2M\times 2M}.
	\label{eq:twin_correlation}
\end{equation}
The twin-representation objective is
\begin{equation}
	\begin{aligned}
		\mathcal L_{\mathrm{twin}}
		&=
		\frac{1}{2M}
		\sum_{d=1}^{2M}
		\bigl(1-C_{12,dd}\bigr)^2
		\\
		&\quad+
		\frac{\lambda_{\mathrm{off}}}{2M(2M-1)}
		\sum_{\substack{d,\ell=1\\d\neq\ell}}^{2M}
		C_{12,d\ell}^{2},
	\end{aligned}
	\label{eq:twin_loss}
\end{equation}
where $d$ and $\ell$ index the $2M$ real-valued latent coordinates and $\lambda_{\mathrm{off}}\geq0$ weights the off-diagonal redundancy penalty. The diagonal term aligns corresponding coordinates across branches, whereas the off-diagonal term discourages cross-coordinate correlation between the two branch representations. Thus, the objective preserves coordinate compatibility for fusion while reducing redundant cross-branch structure. It does not impose orthogonality between the complete latent vectors. Any reduction in adversarial transfer between branches is therefore treated as an empirical consequence of the learned representation diversity rather than as a direct mathematical guarantee of \eqref{eq:twin_loss}.

\subsection{Attack-Aware Sample-Adaptive Latent Masking}
\label{subsec:adaptive_masking}

For branch $i$, a lightweight gating network produces a retention-probability vector
\begin{equation}
	\boldsymbol{\pi}_i
	=
	\pi_{\min}\mathbf 1_M
	+
	(1-\pi_{\min})
	\sigma\!\left(
	\mathcal G_{\vartheta_i}(\mathbf v_i)
	\right)
	\in[\pi_{\min},1]^M,
	\label{eq:retention_probabilities}
\end{equation}
where $\mathcal G_{\vartheta_i}:\mathbb R^{2M}\rightarrow\mathbb R^M$ is the branch-$i$ gating network parameterized by $\vartheta_i$, $\sigma(\cdot)$ is the logistic sigmoid, $\mathbf 1_M$ is the all-ones vector, and $0<\pi_{\min}<1$ prevents vanishing retention probabilities. During training, a binary mask
$\mathbf m_i\in\{0,1\}^M$
is sampled from the Bernoulli probabilities in \eqref{eq:retention_probabilities}; a straight-through binary-Concrete (Gumbel--sigmoid) relaxation is used to propagate gradients through the sampling operation.

Let $\mathcal L_{\mathrm{adv}}$ denote the adversarial objective used to construct training perturbations. The normalized sensitivity of the $k$th complex latent component, evaluated with respect to the pre-mask latent vector, is
\begin{equation}
	\chi_{i,k}
	=
	\frac{
		\left\|
		\nabla_{[\operatorname{Re}(z_{i,k}),\operatorname{Im}(z_{i,k})]}
		\mathcal L_{\mathrm{adv}}
		\right\|_2
	}{
		\displaystyle
		\sum_{\ell=1}^{M}
		\left\|
		\nabla_{[\operatorname{Re}(z_{i,\ell}),\operatorname{Im}(z_{i,\ell})]}
		\mathcal L_{\mathrm{adv}}
		\right\|_2
		+\varepsilon_0
	},
	\label{eq:latent_sensitivity}
\end{equation}
where $\varepsilon_0>0$ ensures numerical stability. The sensitivity scores are detached from the computational graph when the gating network is optimized, thereby avoiding second-order differentiation. Because the scores in \eqref{eq:latent_sensitivity} are already normalized within each branch, the masking objective is defined as
\begin{equation}
	\begin{aligned}
		\mathcal L_{\mathrm{mask}}
		&=
		\frac{1}{2}
		\sum_{i=1}^{2}
		\sum_{k=1}^{M}
		\pi_{i,k}\chi_{i,k}
		\\
		&\quad+
		\frac{\lambda_{\mathrm{bud}}}{2}
		\sum_{i=1}^{2}
		\left(
		\frac{1}{M}
		\sum_{k=1}^{M}\pi_{i,k}
		-q
		\right)^2,
	\end{aligned}
	\label{eq:masking_objective}
\end{equation}
where $q\in[\pi_{\min},1]$ is the target retention ratio and $\lambda_{\mathrm{bud}}\geq0$. The first term penalizes the expected retention of sensitive components, while the second discourages deviation from the prescribed average retention budget.

The masked latent vector and normalized channel input are
\begin{subequations}
	\label{eq:masked_transmission}
	\begin{align}
		\bar{\mathbf z}_i
		&=
		\mathbf m_i\odot\mathbf z_i,
		\label{eq:masked_latent}
		\\
		\mathbf s_i
		&=
		\sqrt{MP_t}\,
		\frac{\bar{\mathbf z}_i}
		{\sqrt{\|\bar{\mathbf z}_i\|_2^2+\varepsilon_0}},
		\qquad i\in\{1,2\}.
		\label{eq:channel_input}
	\end{align}
\end{subequations}
Here, $\odot$ denotes element-wise multiplication. Consequently,
$\|\mathbf s_i\|_2^2/M\leq P_t$
for every realization. At inference,
$K_q=\max\{1,\operatorname{round}(qM)\}$ denotes the number of retained components; the $K_q$ components with the largest entries of $\boldsymbol{\pi}_i$ are kept. Hence, inference requires neither an attack label nor an adversarial gradient.

\subsection{Confidence-Aware Latent Fusion}
\label{subsec:confidence_fusion}

The receiver estimates the instantaneous branch SNR as
\begin{equation}
	\widehat\gamma_i
	=
	\frac{|\widehat h_i|^2P_t}{\sigma_n^2}.
	\label{eq:estimated_branch_snr}
\end{equation}
To obtain a bounded input for the subsequent networks, the SNR in decibels is normalized using fixed bounds from the training SNR range:
\begin{equation}
	\bar\gamma_i
	=
	\operatorname{clip}\!\left(
	\frac{10\log_{10}(\widehat\gamma_i)-\gamma_{\min}^{\mathrm{dB}}}
	{\gamma_{\max}^{\mathrm{dB}}-\gamma_{\min}^{\mathrm{dB}}},
	0,1
	\right).
	\label{eq:normalized_branch_snr}
\end{equation}
Here, $\operatorname{clip}(a,0,1)=\min\{1,\max\{0,a\}\}$, and $\gamma_{\min}^{\mathrm{dB}}$ and $\gamma_{\max}^{\mathrm{dB}}$ are the minimum and maximum SNR values of the training range; they are selected once and fixed at test time.

A confidence network shared by both branches produces reliability logits and normalized fusion weights:
\begin{subequations}
	\label{eq:confidence_fusion}
	\begin{align}
		\ell_i
		&=
		\mathcal Q_{\varphi}\!\left(
		[\mathcal R(\mathbf r_i)^T,\bar\gamma_i]^T
		\right),
		\label{eq:branch_confidence}
		\\
		w_i
		&=
		\frac{\exp(\ell_i/\tau_f)}
		{\displaystyle\sum_{j=1}^{2}\exp(\ell_j/\tau_f)},
		\qquad i\in\{1,2\},
		\label{eq:fusion_weights}
		\\
		\mathbf r_f
		&=
		\sum_{i=1}^{2}w_i\mathbf r_i,
		\label{eq:fused_latent}
	\end{align}
\end{subequations}
where $\mathcal Q_{\varphi}:\mathbb R^{2M+1}\rightarrow\mathbb R$ is the shared confidence network parameterized by $\varphi$, $\ell_i$ is its scalar reliability logit, and $\tau_f>0$ is the fusion temperature. Sharing $\mathcal Q_{\varphi}$ across branches makes the fusion rule invariant to a consistent relabeling of the two branch inputs and allows a branch inferred to be less reliable to receive a smaller weight. Corruption that affects both branches similarly cannot be identified from relative confidence alone and is addressed by the subsequent severity estimator and purifier.

\subsection{Calibrated Blind Corruption-Severity Estimation}
\label{subsec:calibrated_severity}

The raw normalized discrepancy between the two equalized branch observations is
\begin{equation}
	\rho
	=
	\frac{
		\|\mathbf r_1-\mathbf r_2\|_2^2
	}{
		\|\mathbf r_1\|_2^2
		+\|\mathbf r_2\|_2^2
		+\varepsilon_0
	}.
	\label{eq:raw_discrepancy}
\end{equation}
Ordinary channel noise, block-flat Rayleigh fading, and intrinsic branch diversity can also yield nonzero discrepancy. Therefore, clean validation data are used to estimate the SNR-dependent mean and standard deviation of \eqref{eq:raw_discrepancy}. With
$\bar\gamma=(\bar\gamma_1+\bar\gamma_2)/2$,
the calibrated discrepancy is
\begin{equation}
	\rho_{\mathrm{cal}}
	=
	\operatorname{clip}\!\left(
	\frac{
		\rho-\mu_{\rho,0}(\bar\gamma)
	}{
		\tau_\rho\sigma_{\rho,0}(\bar\gamma)+\varepsilon_0
	},
	0,1
	\right),
	\label{eq:calibrated_discrepancy}
\end{equation}
where $\tau_\rho>0$ is a calibration scale factor, while $\mu_{\rho,0}(\cdot)$ and $\sigma_{\rho,0}(\cdot)$ denote the clean-validation mean and standard deviation of $\rho$ as functions of normalized SNR, obtained by interpolation across the training SNR range.

Because the softmax weights are strictly positive, their normalized entropy can be written exactly as
\begin{equation}
	\mathcal H_w
	=
	-\frac{1}{\log 2}
	\sum_{i=1}^{2}w_i\log w_i
	\in[0,1].
	\label{eq:fusion_entropy}
\end{equation}
A blind estimator maps receiver-observable statistics to a continuous corruption-severity score:
\begin{equation}
	\widehat\kappa
	=
	\sigma\!\left(
	\mathcal A_{\zeta}\!\left(
	[\rho_{\mathrm{cal}},\bar\gamma,\mathcal H_w]^T
	\right)
	\right)
	\in(0,1).
	\label{eq:blind_severity}
\end{equation}
Here, $\mathcal A_{\zeta}:\mathbb R^3\rightarrow\mathbb R$ is the severity-estimation network parameterized by $\zeta$; the outer sigmoid maps its output to $(0,1)$. The term \emph{blind} means that the estimator is not supplied with the attack family, perturbation budget, attacked branch, or adversarial channel parameters. The calibrated discrepancy measures abnormal cross-branch disagreement relative to clean behavior at the observed SNR, whereas $\mathcal H_w$ quantifies uncertainty in the branch-reliability assignment. The SNR feature assists in separating corruption-induced discrepancies from those expected under ordinary channel degradation; it does not by itself constitute an attack detector.

For training, each corrupted fused latent vector is paired with a clean reference generated from the same source sample and legitimate channel realization. Let $\mathbf r_f^{\mathrm{corr}}$ and $\mathbf r_f^{\mathrm{clean}}$ denote these corrupted and clean fused latent vectors, respectively. Define the relative latent distortion
\begin{equation}
	d_{\mathrm{rel}}
	=
	\frac{
		\|\mathbf r_f^{\mathrm{corr}}-\mathbf r_f^{\mathrm{clean}}\|_2^2
	}{
		\|\mathbf r_f^{\mathrm{clean}}\|_2^2+\varepsilon_0
	}.
	\label{eq:relative_latent_distortion}
\end{equation}
The regression target and severity loss are
\begin{subequations}
	\label{eq:severity_training}
	\begin{align}
		\kappa
		&=
		\operatorname{clip}\!\left(
		\frac{d_{\mathrm{rel}}}{\kappa_{\max}},
		0,1
		\right),
		\label{eq:severity_target}
		\\
		\mathcal L_{\mathrm{sev}}
		&=
		\mathbb E\!\left[
		(\widehat\kappa-\kappa)^2
		\right],
		\label{eq:severity_loss}
	\end{align}
\end{subequations}
where $\kappa_{\max}>0$ is the relative-distortion level mapped to unit severity and is selected on the validation set. The clean reference is used only during training and calibration and is unavailable at inference.

\subsection{SNR- and Severity-Conditioned Adaptive Diffusion Purification}
\label{subsec:adaptive_diffusion}

Let
$\mathbf u^{\mathrm{obs}}=\mathcal R(\mathbf r_f)$.
The DDIM starting depth is selected as
\begin{equation}
	\begin{aligned}
		t_{\mathrm{start}}
		&=
		t_{\min}
		+
		\operatorname{round}\!\Bigl(
		(t_{\max}-t_{\min})
		\\
		&\qquad\times
		\operatorname{clip}\!\left(
		\lambda_\kappa\widehat\kappa
		+\lambda_\gamma(1-\bar\gamma),
		0,1
		\right)
		\Bigr),
	\end{aligned}
	\label{eq:adaptive_depth}
\end{equation}
where $T$ is the total number of diffusion time steps, $0\leq t_{\min}\leq t_{\max}\leq T$ define the admissible starting-depth range, and $\lambda_\kappa,\lambda_\gamma\geq0$ are severity and SNR weights satisfying $\lambda_\kappa+\lambda_\gamma=1$.
Thus, higher estimated corruption or lower SNR leads to a deeper starting point within the admissible range.

The fused observation is moved to the selected diffusion level according to
\begin{equation}
\begin{aligned}
	\mathbf u_{t_{\mathrm{start}}}
	&=
	\sqrt{\bar\alpha_{t_{\mathrm{start}}}}\,
	\mathbf u^{\mathrm{obs}}
	+
	\sqrt{1-\bar\alpha_{t_{\mathrm{start}}}}\,
	\boldsymbol{\epsilon}^{(0)},\\
	\boldsymbol{\epsilon}^{(0)}
	&\sim\mathcal N(\mathbf 0,\mathbf I_{2M}).
\end{aligned}
	\label{eq:forward_noising}
\end{equation}
where $\{\beta_t\}_{t=1}^{T}$ is the prescribed diffusion noise-variance schedule, $\alpha_t=1-\beta_t$, $\bar\alpha_t=\prod_{s=1}^{t}\alpha_s$, $\bar\alpha_0=1$, and $\mathbf I_{2M}$ is the $2M\times2M$ identity matrix. If $t_{\mathrm{start}}=0$, no forward-noising or reverse-denoising step is applied. The noise-prediction network is conditioned on
\begin{equation}
	\mathbf c
	=
	[\widehat\kappa,\bar\gamma]^T.
	\label{eq:diffusion_condition}
\end{equation}

To cover both consecutive and strided deterministic DDIM sampling, let
$\mathcal T=\{\tau_L,\tau_{L-1},\ldots,\tau_0\}$
be a strictly decreasing sequence satisfying
$\tau_L=t_{\mathrm{start}}$
and
$\tau_0=0$.
Here, $L$ is the number of reverse DDIM network evaluations and $\boldsymbol{\epsilon}_{\phi}(\cdot)$ denotes the conditional noise-prediction network parameterized by $\phi$. For $j=L,L-1,\ldots,1$, the reverse updates are
\begin{subequations}
	\label{eq:ddim_purification}
	\begin{align}
		\widehat{\boldsymbol{\epsilon}}_{\tau_j}
		&=
		\boldsymbol{\epsilon}_{\phi}
		(\mathbf u_{\tau_j},\tau_j,\mathbf c),
		\label{eq:predicted_noise}
		\\
		\widehat{\mathbf u}_{0}^{(\tau_j)}
		&=
		\frac{
			\mathbf u_{\tau_j}
			-
			\sqrt{1-\bar\alpha_{\tau_j}}\,
			\widehat{\boldsymbol{\epsilon}}_{\tau_j}
		}{
			\sqrt{\bar\alpha_{\tau_j}}
		},
		\label{eq:estimated_clean_latent}
		\\
		\mathbf u_{\tau_{j-1}}
		&=
		\sqrt{\bar\alpha_{\tau_{j-1}}}\,
		\widehat{\mathbf u}_{0}^{(\tau_j)}
		+
		\sqrt{1-\bar\alpha_{\tau_{j-1}}}\,
		\widehat{\boldsymbol{\epsilon}}_{\tau_j}.
		\label{eq:ddim_update}
	\end{align}
\end{subequations}
In \eqref{eq:ddim_purification}, $\widehat{\boldsymbol{\epsilon}}_{\tau_j}$ is the predicted noise and $\widehat{\mathbf u}_{0}^{(\tau_j)}$ is the corresponding estimate of the clean real-valued latent. Equation~\eqref{eq:ddim_update} is the deterministic DDIM update, with the DDIM stochasticity parameter set to $\eta=0$. Choosing every integer time index recovers a consecutive schedule, whereas selecting a subset of indices reduces the number $L$ of diffusion-network evaluations.

Let $g_{\psi}$ denote the semantic image decoder parameterized by $\psi$. The purified latent vector and reconstructed image are
\begin{subequations}
	\label{eq:semantic_reconstruction}
	\begin{align}
		\widehat{\mathbf z}
		&=
		\mathcal R^{-1}(\mathbf u_{\tau_0}),
		\label{eq:purified_latent}
		\\
		\widehat{\mathbf x}
		&=
		g_{\psi}(\widehat{\mathbf z}).
		\label{eq:reconstructed_image}
	\end{align}
\end{subequations}

\subsection{Joint Training Objective}
\label{subsec:joint_training}

A classifier $p_{\nu}(c\mid\cdot)$, parameterized by $\nu$, is pretrained on clean CIFAR-100 images and remains fixed while the communication system is optimized. Let $c^{\star}$ denote the ground-truth class. The reconstruction and semantic-task losses are
\begin{subequations}
	\label{eq:principal_losses}
	\begin{align}
		\mathcal L_{\mathrm{rec}}
		&=
		\mathbb E\!\left[
		\frac{\|\mathbf x-\widehat{\mathbf x}\|_2^2}{HWC}
		\right],
		\label{eq:reconstruction_loss}
		\\
		\mathcal L_{\mathrm{sem}}
		&=
		\mathbb E\!\left[
		-\log p_{\nu}(c^{\star}\mid\widehat{\mathbf x})
		\right].
		\label{eq:semantic_loss}
	\end{align}
\end{subequations}
The target in \eqref{eq:reconstruction_loss} is the clean source image even when the transmitted input or channel observation is adversarially corrupted.

For diffusion training, let
$\mathbf u_0^{\mathrm{clean}}=\mathcal R(\mathbf r_f^{\mathrm{clean}})$
be the clean fused latent associated with a paired corrupted observation. A diffusion time $t$ is sampled uniformly from $\{1,\ldots,T\}$, i.e., $t\sim\mathcal U\{1,\ldots,T\}$, and noise
$\boldsymbol{\epsilon}\sim\mathcal N(\mathbf0,\mathbf I_{2M})$
are sampled, and
\begin{equation}
	\mathbf u_t^{\mathrm{clean}}
	=
	\sqrt{\bar\alpha_t}\,
	\mathbf u_0^{\mathrm{clean}}
	+
	\sqrt{1-\bar\alpha_t}\,
	\boldsymbol{\epsilon}.
	\label{eq:diffusion_training_sample}
\end{equation}
The conditional noise-prediction loss is
\begin{equation}
	\mathcal L_{\mathrm{diff}}
	=
	\mathbb E_{\mathbf u_0^{\mathrm{clean}},t,\boldsymbol{\epsilon},\mathbf c}
	\!\left[
	\frac{1}{2M}
	\left\|
	\boldsymbol{\epsilon}
	-
	\boldsymbol{\epsilon}_{\phi}
	(\mathbf u_t^{\mathrm{clean}},t,\mathbf c)
	\right\|_2^2
	\right],
	\label{eq:diffusion_loss}
\end{equation}
where the condition $\mathbf c$ is computed from the paired corrupted observation. This formulation trains the diffusion network toward the clean latent distribution while allowing its denoising behavior to depend on the inferred corruption severity and channel quality.

The adversarial objective used to generate source- or channel-domain training perturbations and to compute \eqref{eq:latent_sensitivity} is
\begin{equation}
	\mathcal L_{\mathrm{adv}}
	=
	\mathcal L_{\mathrm{rec}}
	+
	\lambda_{\mathrm{adv,sem}}
	\mathcal L_{\mathrm{sem}},
	\label{eq:adversarial_objective}
\end{equation}
where $\lambda_{\mathrm{adv,sem}}\geq0$ controls the semantic-task contribution to attack generation. The objective is maximized with respect to the admissible source perturbation or adversarial waveform while the model parameters are held fixed during attack generation.

The complete minimization objective is
\begin{equation}
	\begin{aligned}
		\mathcal L_{\mathrm{total}}
		&=
		\mathcal L_{\mathrm{rec}}
		+
		\lambda_{\mathrm{sem}}\mathcal L_{\mathrm{sem}}
		+
		\lambda_{\mathrm{twin}}\mathcal L_{\mathrm{twin}}
		+
		\lambda_{\mathrm{mask}}\mathcal L_{\mathrm{mask}}
		\\
		&\quad+
		\lambda_{\mathrm{sev}}\mathcal L_{\mathrm{sev}}
		+
		\lambda_{\mathrm{diff}}\mathcal L_{\mathrm{diff}},
	\end{aligned}
	\label{eq:total_objective}
\end{equation}
where $\lambda_{\mathrm{sem}},\lambda_{\mathrm{twin}},\lambda_{\mathrm{mask}},\lambda_{\mathrm{sev}},$ and $\lambda_{\mathrm{diff}}$ are nonnegative loss weights. Hyperparameters, SNR-normalization bounds, calibration statistics, and severity and diffusion-depth constants are selected using the validation set and are then fixed for all test attacks, SNRs, and channel models.

\subsection{Training and Inference Procedure}
\label{subsec:training_inference}

Training proceeds in four stages. First, the downstream classifier is pretrained on clean CIFAR-100 images and frozen. Second, the twin encoders, projection heads, gating networks, and decoder are trained using a mixture of benign and adversarially corrupted examples. Third, the confidence-fusion network, severity estimator, SNR-dependent calibration statistics, and conditional diffusion model are trained from paired clean and corrupted latent observations. Finally, all differentiable communication and defense modules are jointly fine-tuned using \eqref{eq:total_objective}, while the classifier remains frozen.

At inference, the transmitter computes the two latent representations, applies deterministic top-$K_q$ masking, normalizes the transmitted vectors, and sends them over orthogonal resources. The receiver requires only the legitimate-channel estimate and nominal noise variance; no attack metadata or clean reference is used. The dominant per-sample computational cost is
\begin{equation}
	\mathcal O\!\left(
	2C_{\mathrm{enc}}
	+2C_{\mathrm{gate}}
	+C_{\mathrm{fus}}
	+C_{\mathrm{sev}}
	+LC_{\mathrm{diff}}
	+C_{\mathrm{dec}}
	\right),
	\label{eq:inference_complexity}
\end{equation}
where $C_{\mathrm{enc}},C_{\mathrm{gate}},C_{\mathrm{fus}},C_{\mathrm{sev}},C_{\mathrm{diff}},$ and $C_{\mathrm{dec}}$ denote the costs of one encoder--projection branch, one gating network, confidence-aware fusion, severity estimation, one diffusion-network evaluation, and semantic decoding, respectively; $L$ is the number of DDIM network evaluations selected by the adaptive schedule. Higher-SNR or lower-severity observations generally invoke fewer reverse evaluations than more severely corrupted observations.

\section{Simulation Results}
\label{sec:simulation_results}

\subsection{Experimental Setup}
\label{subsec:experimental_setup}

The proposed framework is evaluated on the CIFAR-100 dataset, which contains
$50{,}000$ training images and $10{,}000$ test images of size
$32\times32\times3$ distributed across $100$ object classes. A validation
subset is used for hyperparameter selection, SNR-dependent
clean-discrepancy calibration, checkpoint selection, and adjustment of the
adaptive purification parameters. Reconstruction fidelity is quantified
using PSNR, whereas task-level semantic
preservation is evaluated using the Top-1 accuracy of a fixed CIFAR-100
classifier.

For images normalized to $[0,1]$, the mean-squared error (MSE) and PSNR are
computed as
\begin{equation}
	\mathrm{MSE}
	=
	\frac{
		\|\mathbf{x}-\widehat{\mathbf{x}}\|_2^2
	}{
		HWC
	},
	\qquad
	\mathrm{PSNR}
	=
	10\log_{10}
	\left(
	\frac{1}{\mathrm{MSE}}
	\right)
	\ \mathrm{dB}.
	\label{eq:evaluation_psnr}
\end{equation}

All compared schemes operate under the same aggregate budget of $K$
complex channel uses. TwinViT-DeepJSCC allocates $M=K/2$ mutually
orthogonal channel uses to each branch and therefore does not receive
additional bandwidth relative to the single-branch baselines. The nominal
average channel SNR is varied over
$\{-5,0,5,10,15,20\}$~dB. Both additive white Gaussian noise (AWGN) and
block-flat Rayleigh fading channels are considered. Unless otherwise
stated, perfect estimation of the legitimate channel coefficients is
assumed, i.e., $\widehat h_i=h_i$.

The following schemes are used for comparison:

\begin{enumerate}
	\item \emph{ViT-DeepJSCC}: a single-branch ViT-based DeepJSCC
	transceiver without an explicit adversarial defense;
	
	\item \emph{Adversarial Training}: the same single-branch transceiver
	trained using adversarially perturbed source samples;
	
	\item \emph{Diffusion-Only}: a single-branch transceiver equipped
	with receiver-side diffusion purification but without twin
	representation learning, adaptive latent masking, or
	confidence-aware fusion; and
	
	\item \emph{TwinViT-DeepJSCC}: the complete preventive--corrective
	framework proposed in this paper.
\end{enumerate}

The dataset split, aggregate channel-use budget, channel realizations,
attack constraints, and downstream classifier are kept identical across
the compared methods. This evaluation protocol prevents the proposed
method from obtaining an advantage through additional communication
resources or a different semantic classifier.

Source-domain robustness is evaluated using FGSM, 20-step PGD, NES, and
CW attacks. Unless otherwise specified, FGSM and PGD use an
$\ell_{\infty}$ perturbation budget of $\epsilon_x=8/255$. The PGD
sensitivity experiments additionally consider
$\epsilon_x\in\{2/255,4/255,8/255\}$. NES represents a score-based
black-box attack, whereas CW represents an optimization-based attack under
matched perturbation constraints.

Channel-domain experiments consider random Gaussian jamming and a
channel-aware adversarial waveform satisfying
\eqref{eq:channel_attack_constraint}. The attacker strength is expressed
through the nominal transmit-side adversarial-to-signal power ratio
defined in \eqref{eq:asr}. All channel-domain results are obtained through
simulation and should not be interpreted as hardware over-the-air
measurements.

The models are trained for at most $1000$ epochs, and the checkpoint with
the best validation performance is retained. Loss coefficients, target
retention ratio, fusion temperature, discrepancy-calibration constants,
severity-normalization parameters, and diffusion-depth limits are selected
using the validation subset and remain fixed across all test attacks, SNR
values, and channel models.

\subsection{Reconstruction Robustness under Source- and Channel-Domain Attacks}
\label{subsec:reconstruction_robustness}

\begin{figure*}[!t]
    \centering
    \includegraphics[width=0.46\textwidth]{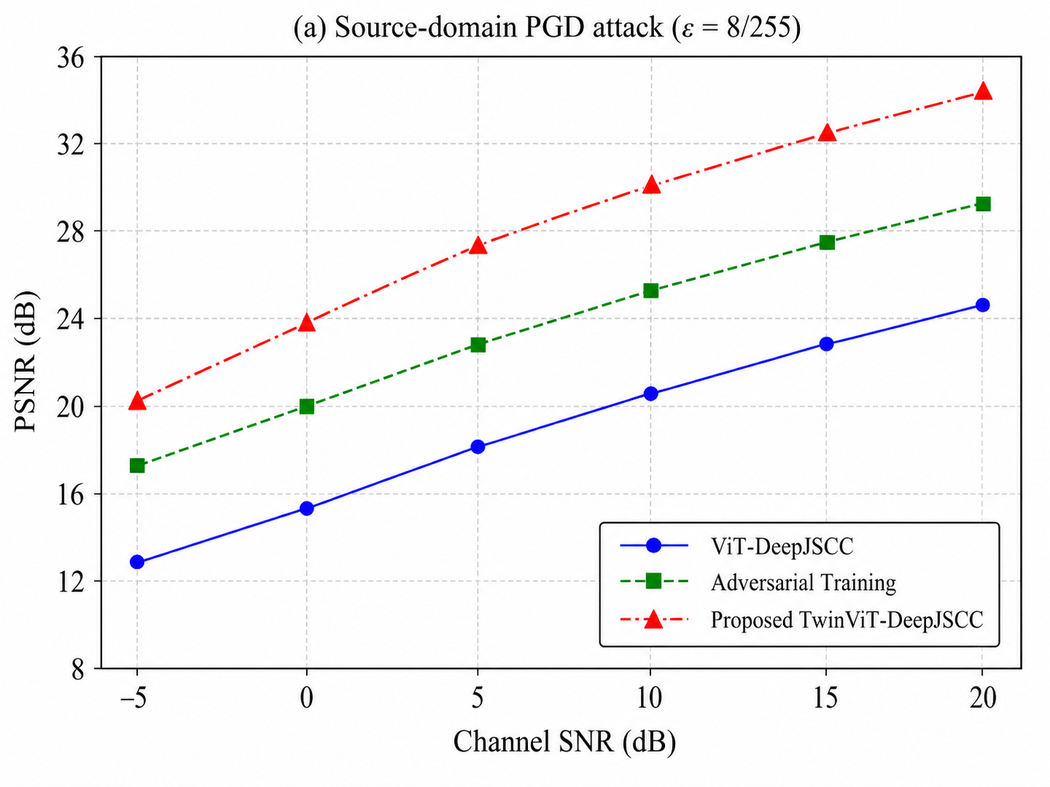}\hfill
    \includegraphics[width=0.46\textwidth]{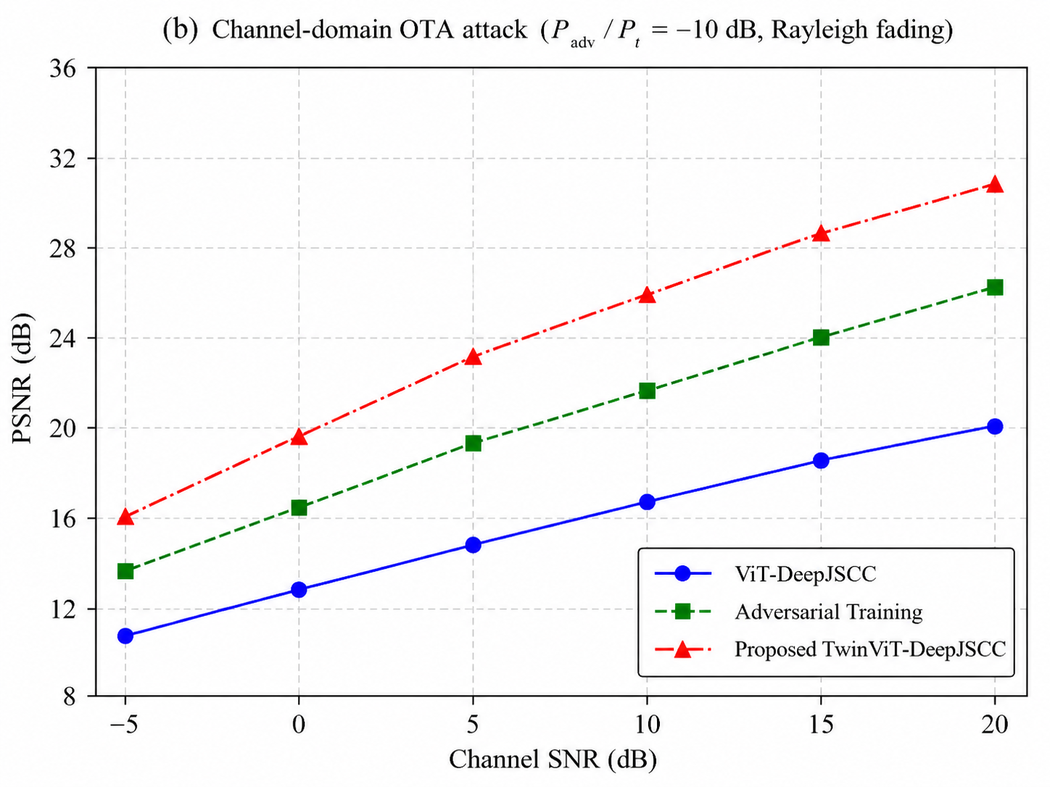}
    \caption{Reconstruction quality versus channel SNR under
        (a) a source-domain 20-step PGD attack with
        $\epsilon_x=8/255$ and
        (b) a simulated channel-aware adversarial waveform over block-flat
        Rayleigh fading with $\mathrm{ASR}=-10$~dB. All methods use the same
        aggregate channel-use budget and attack constraints.}
    \label{fig:fig2_psnr}
\end{figure*}

Fig.~\ref{fig:fig2_psnr} evaluates reconstruction fidelity under two
distinct attack surfaces. In Fig.~\ref{fig:fig2_psnr}(a), a 20-step PGD
perturbation is applied to the source image before semantic encoding. In
Fig.~\ref{fig:fig2_psnr}(b), a channel-aware adversarial waveform is
injected after semantic encoding and directly corrupts the transmitted
latent symbols.

As shown in Fig.~\ref{fig:fig2_psnr}(a), the reconstruction quality of all
methods improves with channel SNR because the relative contribution of
stochastic channel noise decreases. Nevertheless, the undefended
ViT-DeepJSCC baseline remains vulnerable to the source-domain
perturbation, including at relatively high SNR. Adversarial training
substantially improves the baseline because its training distribution
explicitly includes source-side perturbations. TwinViT-DeepJSCC
consistently achieves the highest PSNR over the evaluated SNR range,
demonstrating the benefit of combining twin representation learning,
sensitivity-aware latent masking, and adaptive receiver-side correction.

Fig.~\ref{fig:fig2_psnr}(b) considers an adversarial waveform injected
after the semantic encoder. In this case, source-domain adversarial
training alone provides limited protection because the attack directly
modifies the transmitted latent symbols. The proposed framework retains
the highest PSNR throughout the evaluated SNR range. This result supports
the coordinated use of transmitter-side latent protection and
receiver-side adaptive purification when adversarial and stochastic
channel distortions coexist.

\subsection{Robustness to Diverse Source-Domain Attacks}
\label{subsec:diverse_source_attacks}

\begin{figure*}[!t]
    \centering
    \includegraphics[width=0.43\textwidth]{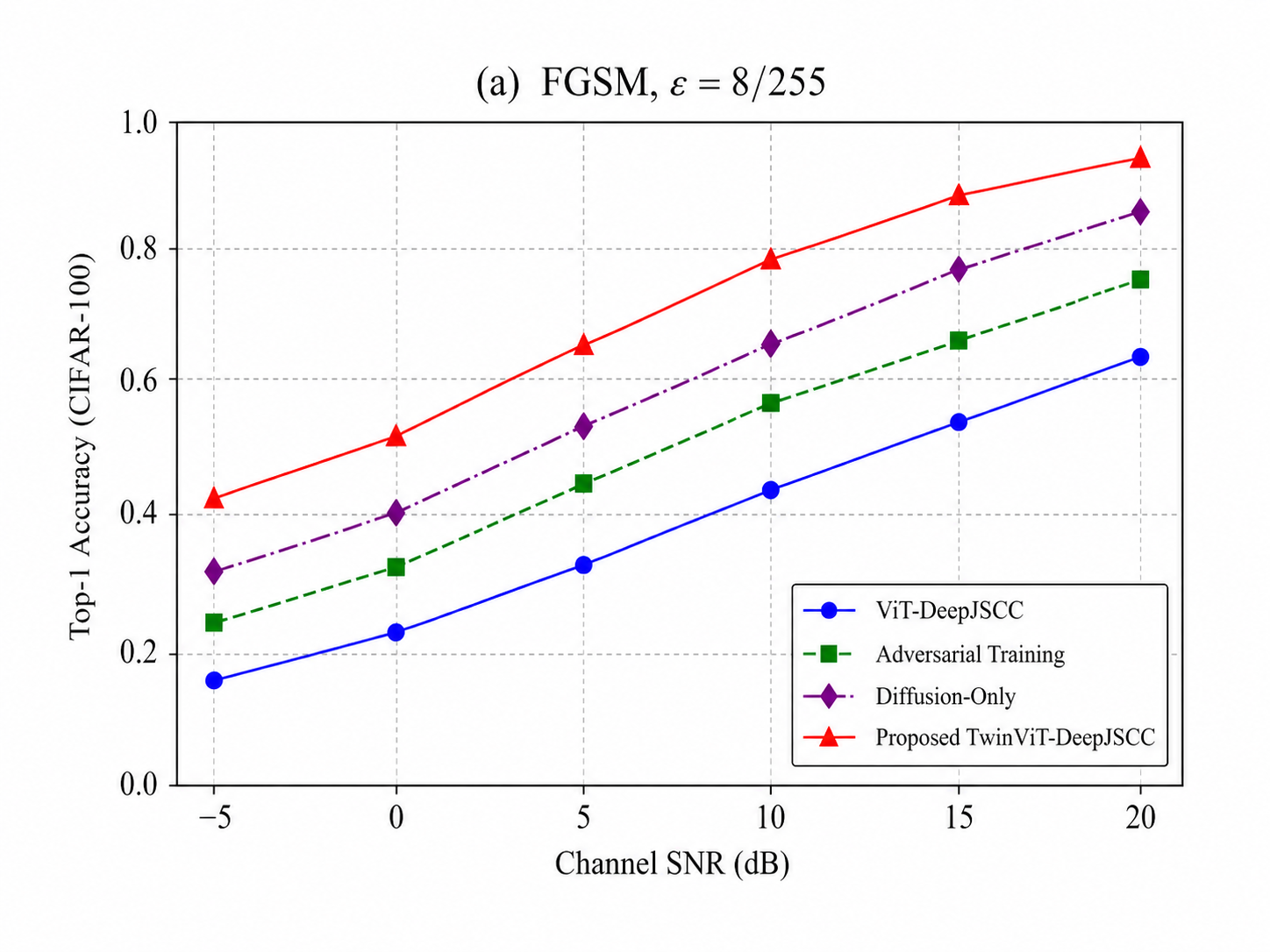}\hfill
    \includegraphics[width=0.43\textwidth]{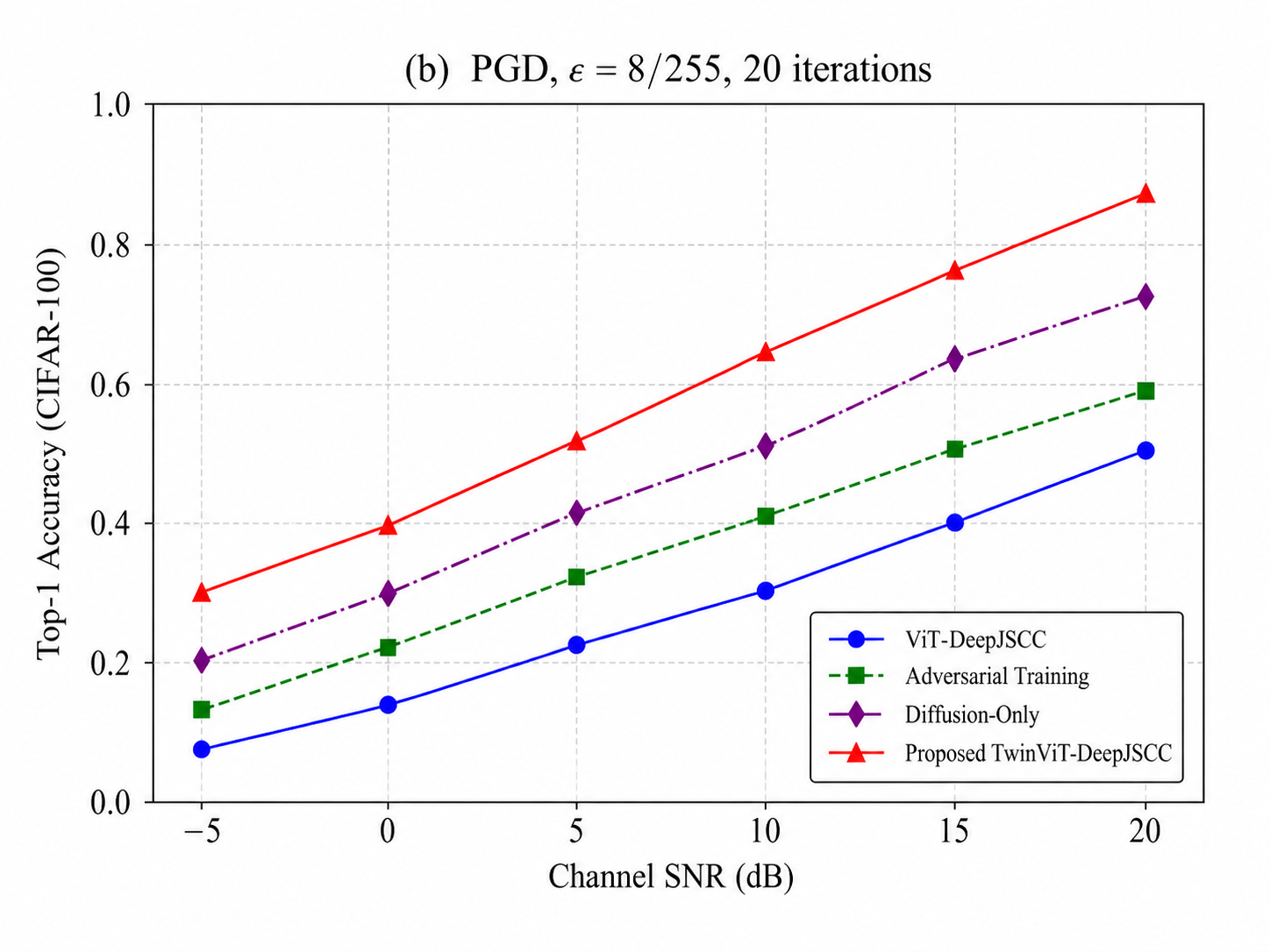}\\[-0.5mm]
    \includegraphics[width=0.43\textwidth]{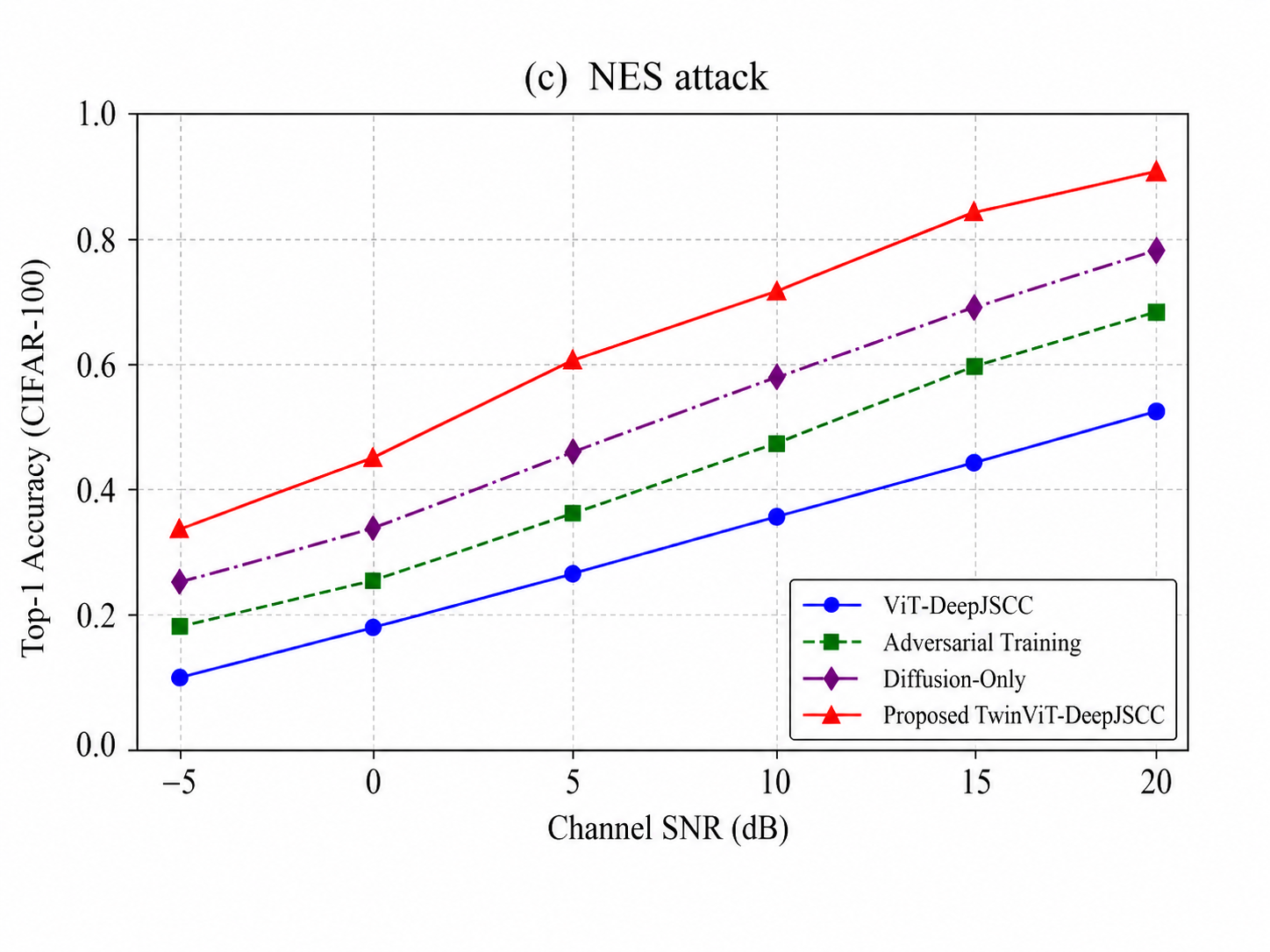}\hfill
    \includegraphics[width=0.43\textwidth]{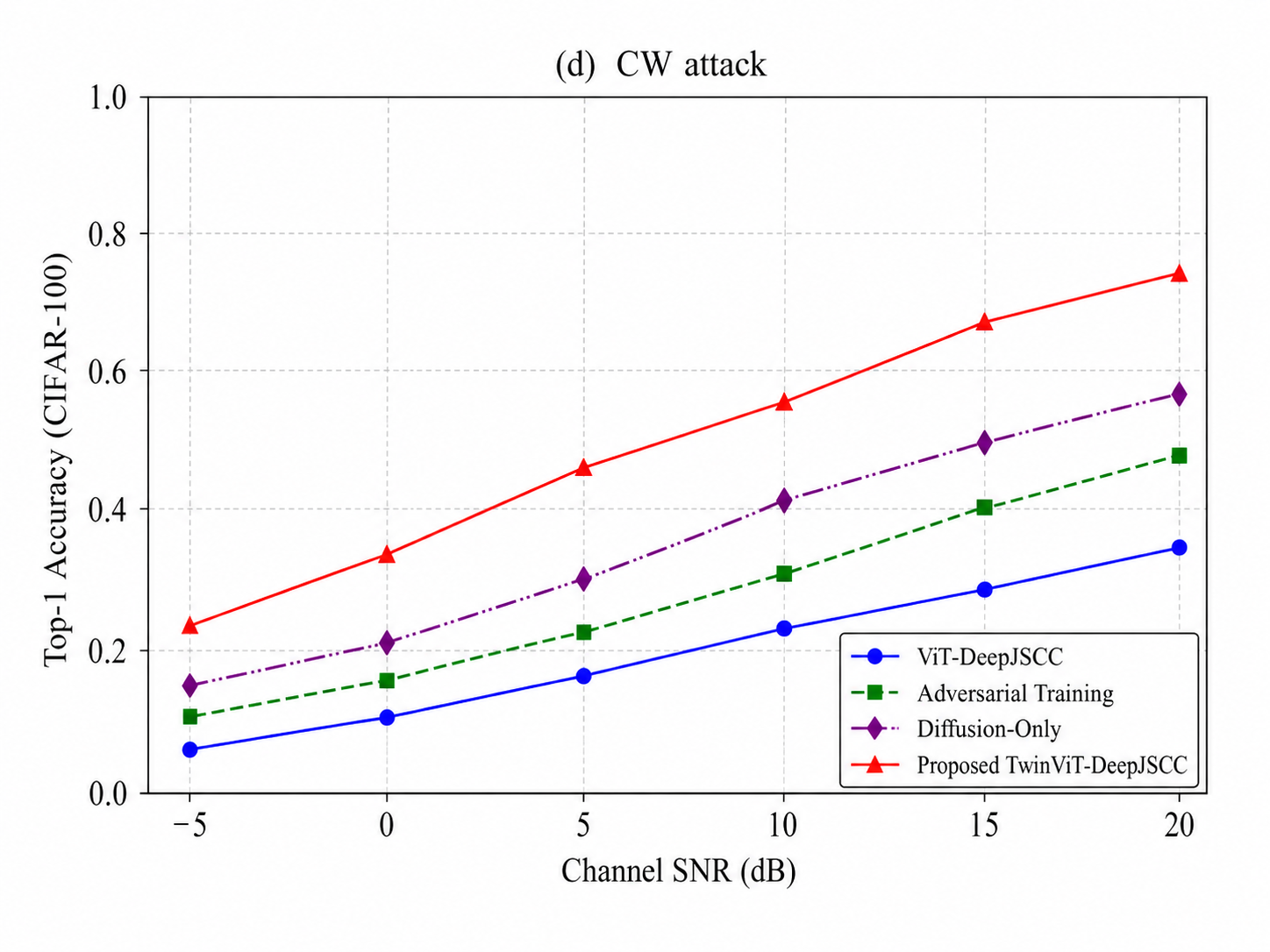}
    \caption{Top-1 semantic accuracy versus channel SNR under
        (a) FGSM,
        (b) 20-step PGD,
        (c) score-based NES, and
        (d) optimization-based CW source-domain attacks.
        FGSM and PGD use $\epsilon_x=8/255$, and NES and CW use matched
        perturbation constraints.}
    \label{fig:fig3_source_attacks}
\end{figure*}

Fig.~\ref{fig:fig3_source_attacks} reports Top-1 semantic accuracy under
FGSM, PGD, NES, and CW source-domain attacks. These attacks respectively
represent a single-step white-box method, an iterative white-box method, a
score-based black-box method, and an optimization-based method.

Under the single-step FGSM attack shown in
Fig.~\ref{fig:fig3_source_attacks}(a), all explicit defense mechanisms
improve upon the undefended ViT-DeepJSCC baseline. TwinViT-DeepJSCC
nevertheless provides the highest semantic accuracy across the full SNR
range, indicating that the proposed protection remains effective even
under a relatively simple gradient-based perturbation.

The stronger iterative PGD attack in
Fig.~\ref{fig:fig3_source_attacks}(b) creates a more pronounced separation
between the proposed method and the competing schemes. Although
adversarial training improves robustness relative to the undefended
baseline, the complete preventive--corrective framework consistently
achieves the highest Top-1 accuracy. This result shows that the reported
gain is not limited to weak single-step attacks.

The performance under the score-based NES attack in
Fig.~\ref{fig:fig3_source_attacks}(c) is particularly important because
the attacker does not require direct access to the exact transceiver
gradients. TwinViT-DeepJSCC maintains the best performance in this
black-box setting, whereas adversarial training and diffusion-only
purification exhibit greater sensitivity to attack mismatch.

Finally, Fig.~\ref{fig:fig3_source_attacks}(d) shows that the proposed
framework also preserves the highest semantic accuracy under the
optimization-based CW attack. Taken together, the four panels of
Fig.~\ref{fig:fig3_source_attacks} demonstrate that the observed
robustness improvement is not tied to a single attack algorithm or access
assumption.

\subsection{Effect of Attack Strength and Channel Model}
\label{subsec:attack_strength}

\begin{figure*}[!t]
    \centering
    \includegraphics[width=0.285\textwidth]{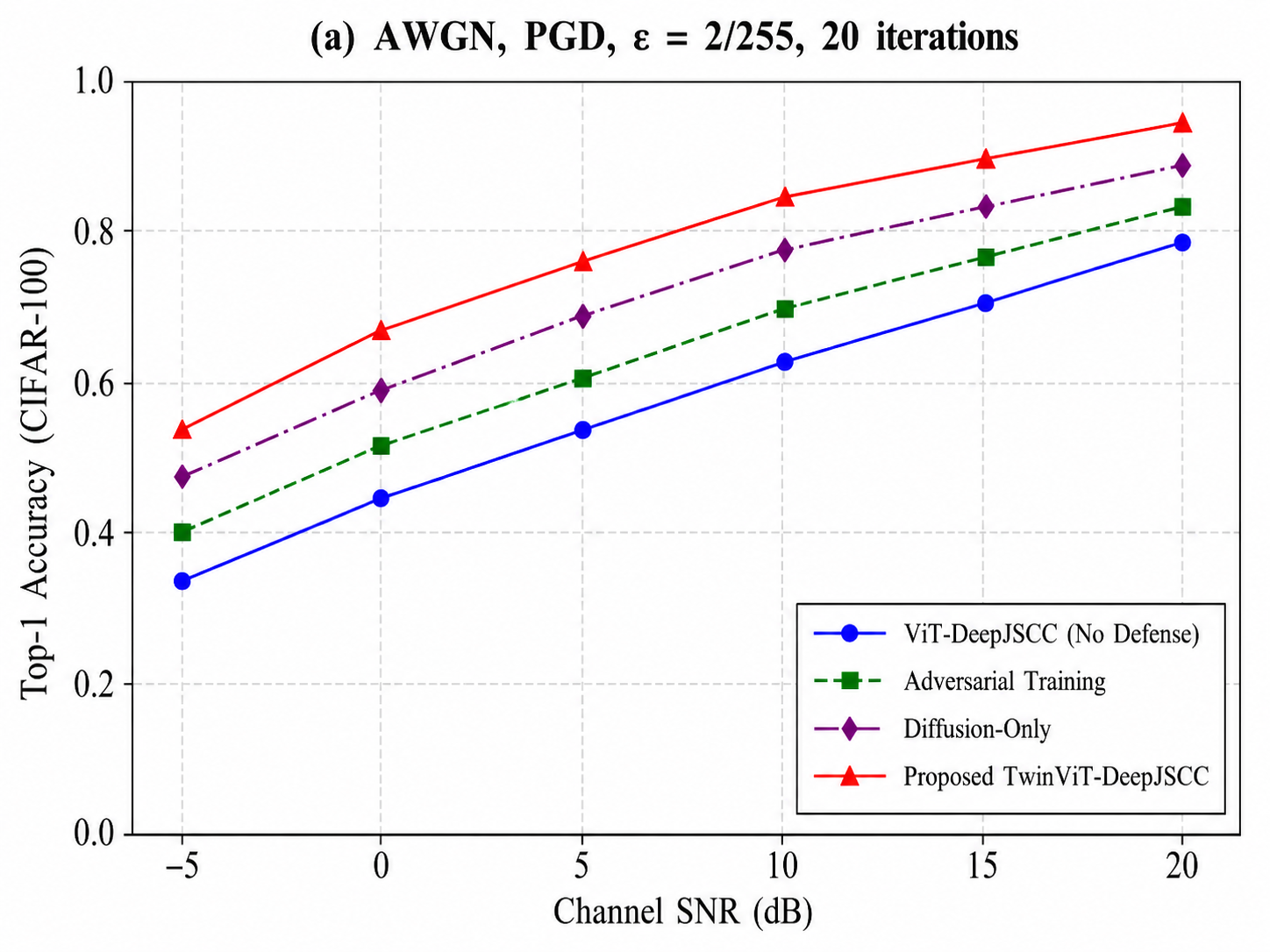}\hfill
    \includegraphics[width=0.285\textwidth]{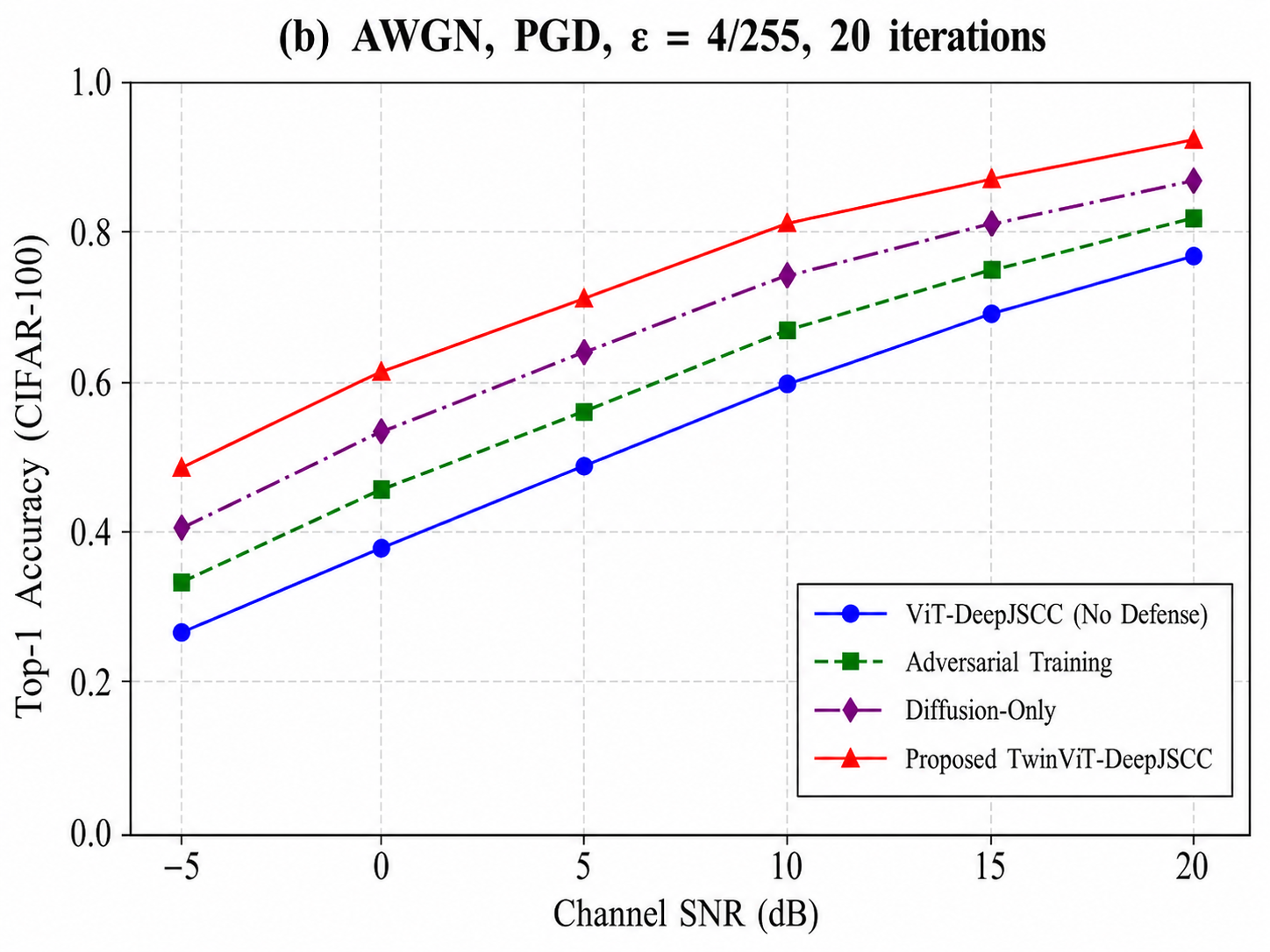}\hfill
    \includegraphics[width=0.285\textwidth]{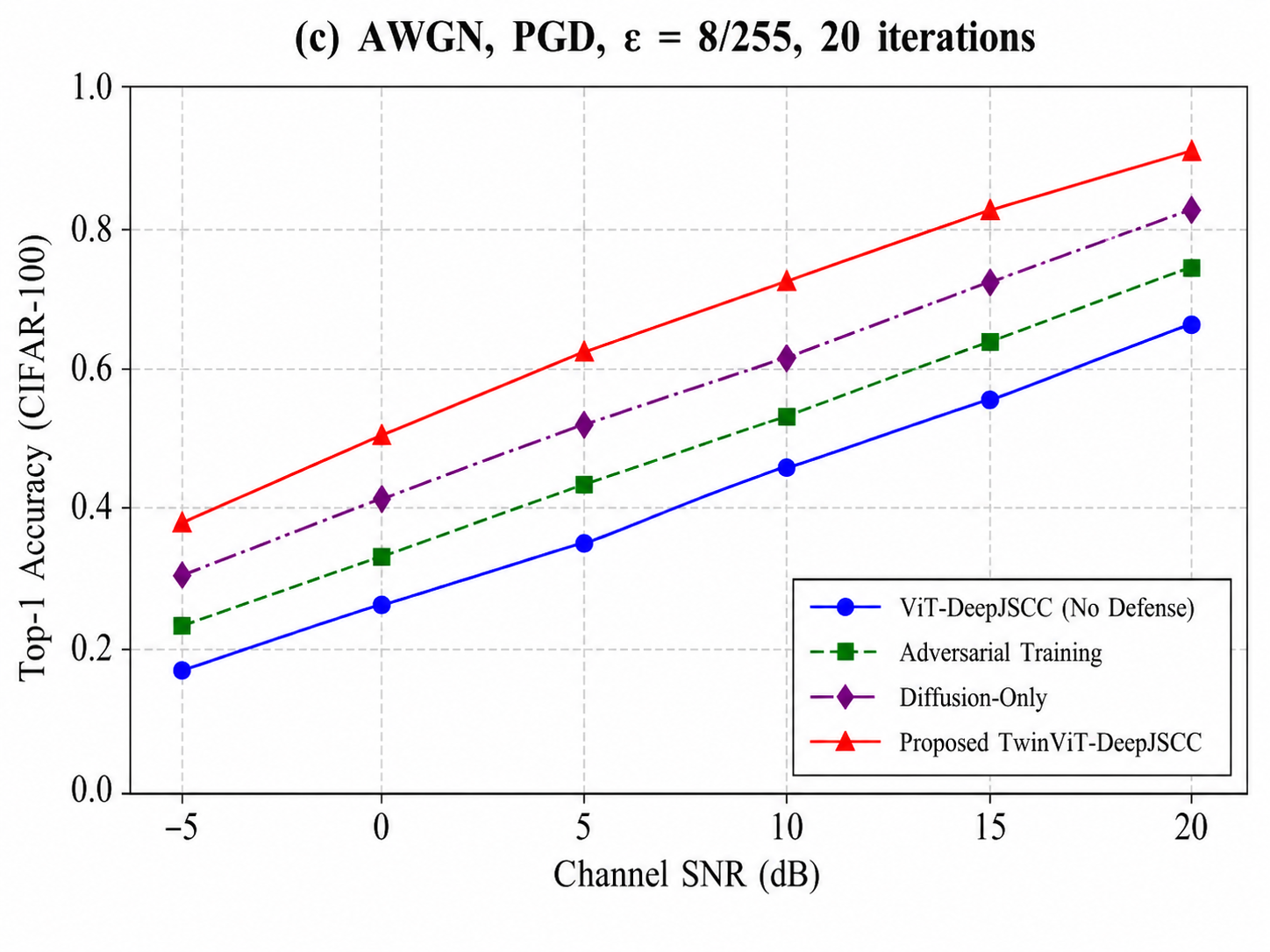}\\[-0.5mm]
    \includegraphics[width=0.285\textwidth]{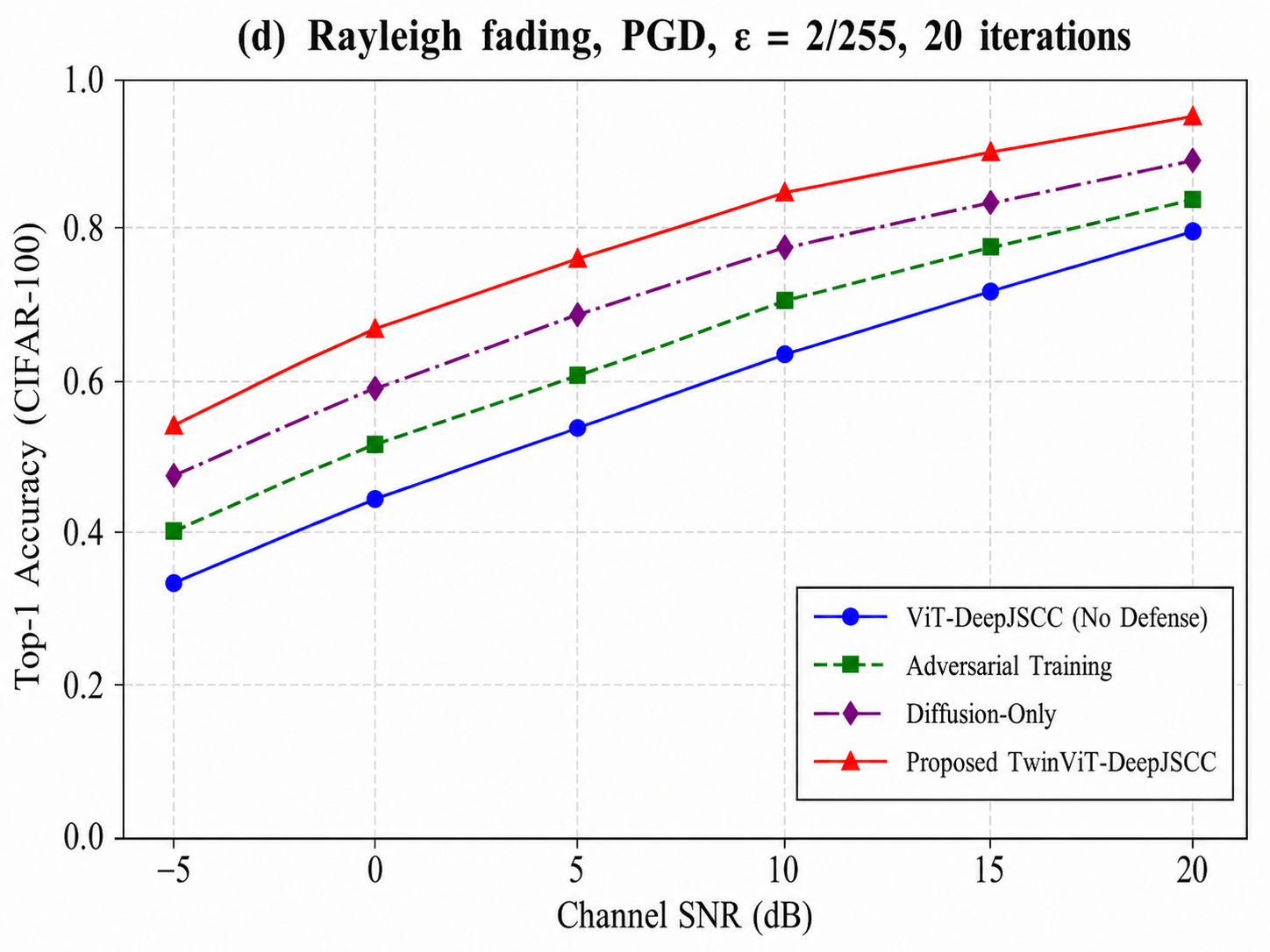}\hfill
    \includegraphics[width=0.285\textwidth]{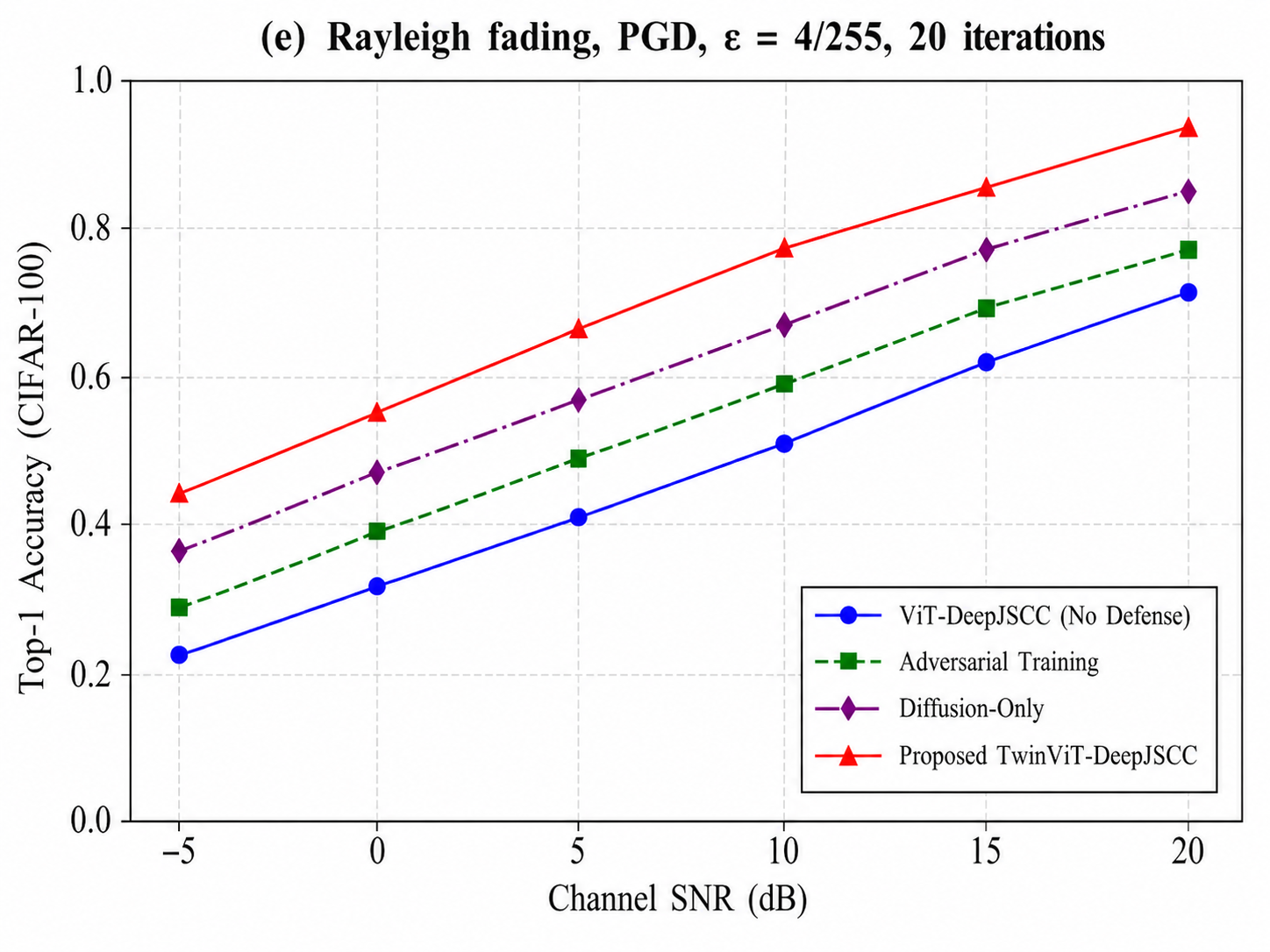}\hfill
    \includegraphics[width=0.285\textwidth]{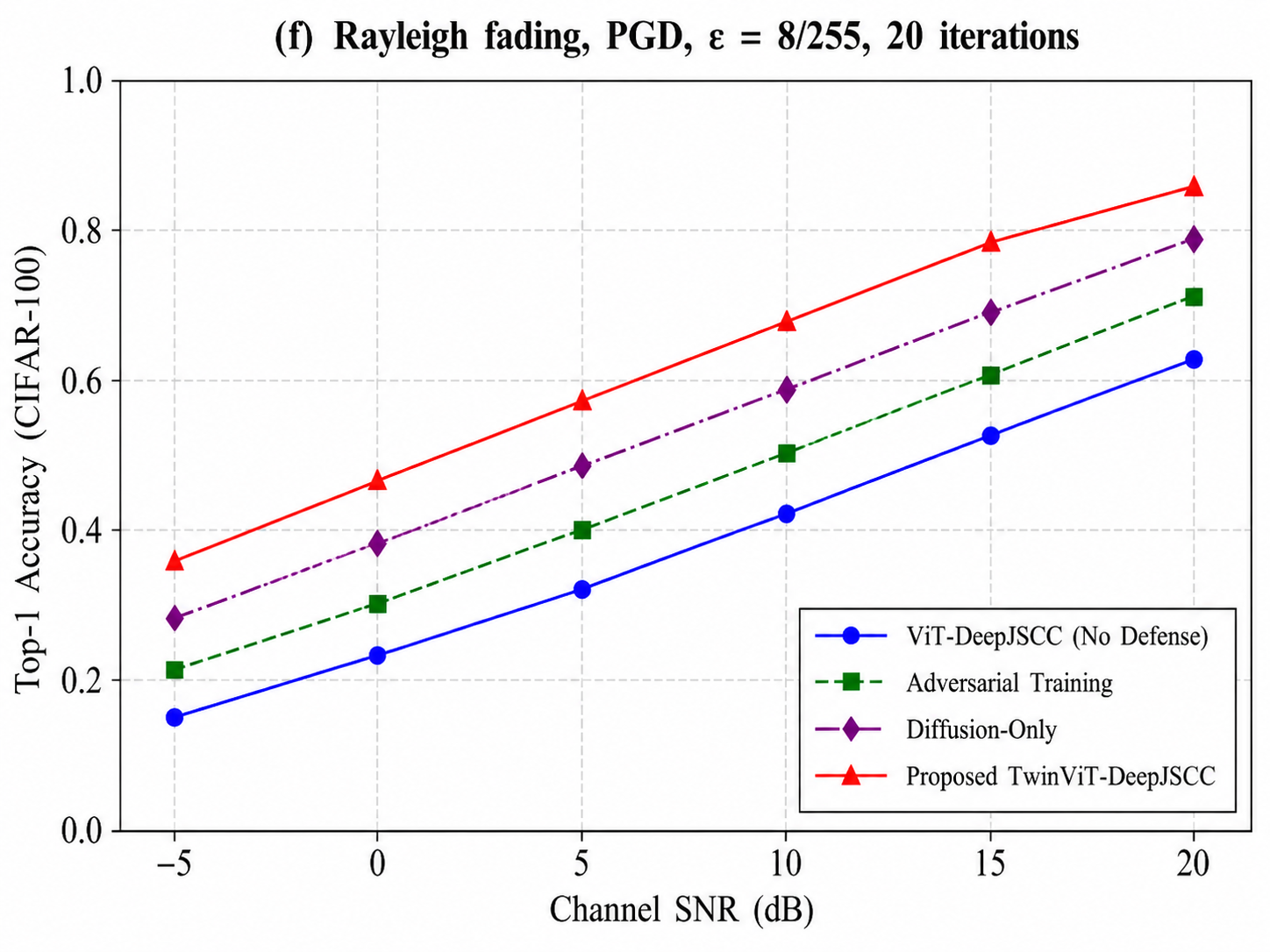}
    \caption{Top-1 semantic accuracy versus channel SNR under 20-step PGD:
        AWGN with (a) $\epsilon_x=2/255$, (b) $\epsilon_x=4/255$, and
        (c) $\epsilon_x=8/255$; block-flat Rayleigh fading with
        (d) $\epsilon_x=2/255$, (e) $\epsilon_x=4/255$, and
        (f) $\epsilon_x=8/255$.}
    \label{fig:fig4_pgd_budgets}
\end{figure*}

Fig.~\ref{fig:fig4_pgd_budgets} evaluates the sensitivity of the compared
schemes to the PGD perturbation budget under AWGN and block-flat Rayleigh
fading. Panels (a)--(c) correspond to AWGN, whereas panels (d)--(f)
correspond to block-flat Rayleigh fading.

Figs.~\ref{fig:fig4_pgd_budgets}(a)--(c) show that semantic accuracy
decreases as the PGD budget increases. The reduction occurs for all
methods because a larger $\epsilon_x$ permits a stronger perturbation of
the source image. TwinViT-DeepJSCC maintains the highest accuracy under
all three AWGN configurations and exhibits a more gradual degradation
than the competing schemes.

The block-flat Rayleigh fading results in
Figs.~\ref{fig:fig4_pgd_budgets}(d)--(f) are generally lower than the
corresponding AWGN results because block-flat Rayleigh fading introduces multiplicative channel variation in addition to
the source-domain perturbation and additive channel noise. Nevertheless,
the proposed framework preserves the strongest semantic accuracy across
all perturbation budgets and SNR values.

The complete set of results in Fig.~\ref{fig:fig4_pgd_budgets} indicates
that TwinViT-DeepJSCC is not tuned to a single nominal attack budget or
channel distribution. Its accuracy decreases as the source perturbation
becomes stronger, as expected, but the degradation remains more gradual
than that of the competing schemes.

\subsection{Robustness against Channel-Domain Adversarial Waveforms}
\label{subsec:channel_results}

\begin{figure*}[!t]
	\centering
	\includegraphics[width=0.92\textwidth]{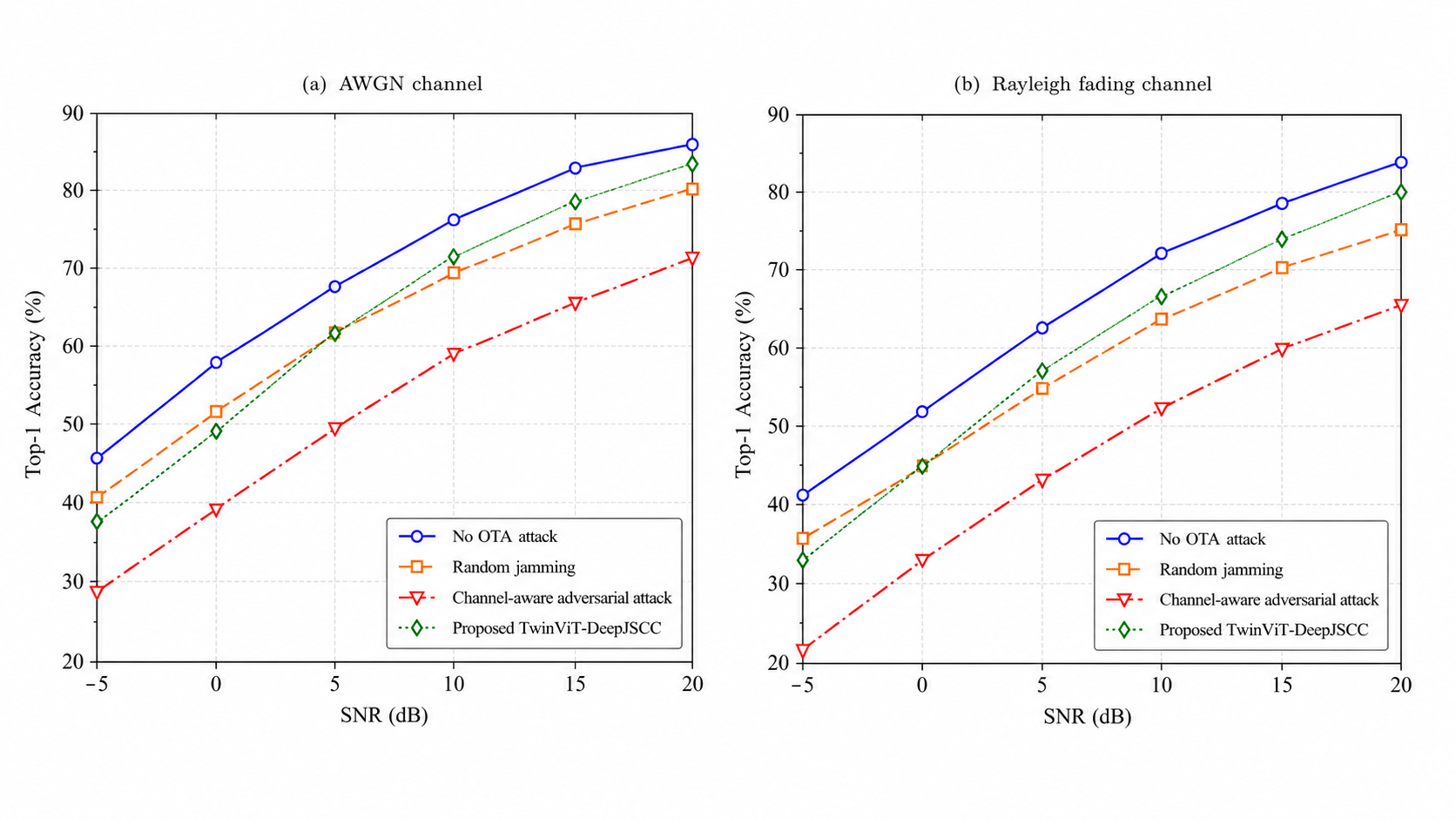}
	\caption{Top-1 semantic accuracy versus channel SNR under
		channel-domain interference over
		(a) AWGN and
		(b) block-flat Rayleigh fading. The benign condition, random Gaussian
		jamming, and channel-aware adversarial waveform are evaluated under
		identical communication and attack-power budgets.}
	\label{fig:fig5_channel}
\end{figure*}

Fig.~\ref{fig:fig5_channel} compares benign transmission, random Gaussian
jamming, a channel-aware adversarial waveform, and the proposed defended
transmission over AWGN and block-flat Rayleigh fading.

Random Gaussian jamming is generally less damaging than the optimized channel-aware waveform because its energy is not aligned with vulnerable semantic directions. At very low SNR, the proposed defended curve may slightly cross the random-jamming curve because strong channel noise reduces the discriminability of the inter-branch discrepancy, confidence, and severity statistics used by the adaptive receiver, thereby limiting the benefit of adaptive fusion and purification. As SNR increases, these statistics become more informative and the proposed receiver provides a clearer advantage. Overall, TwinViT-DeepJSCC recovers a substantial portion of the semantic accuracy lost under channel-domain manipulation while remaining below the benign no-attack condition.

\subsection{Module-Level Ablation Study}
\label{subsec:module_ablation}

\begin{figure*}[!t]
	\centering
	\includegraphics[width=0.92\textwidth]{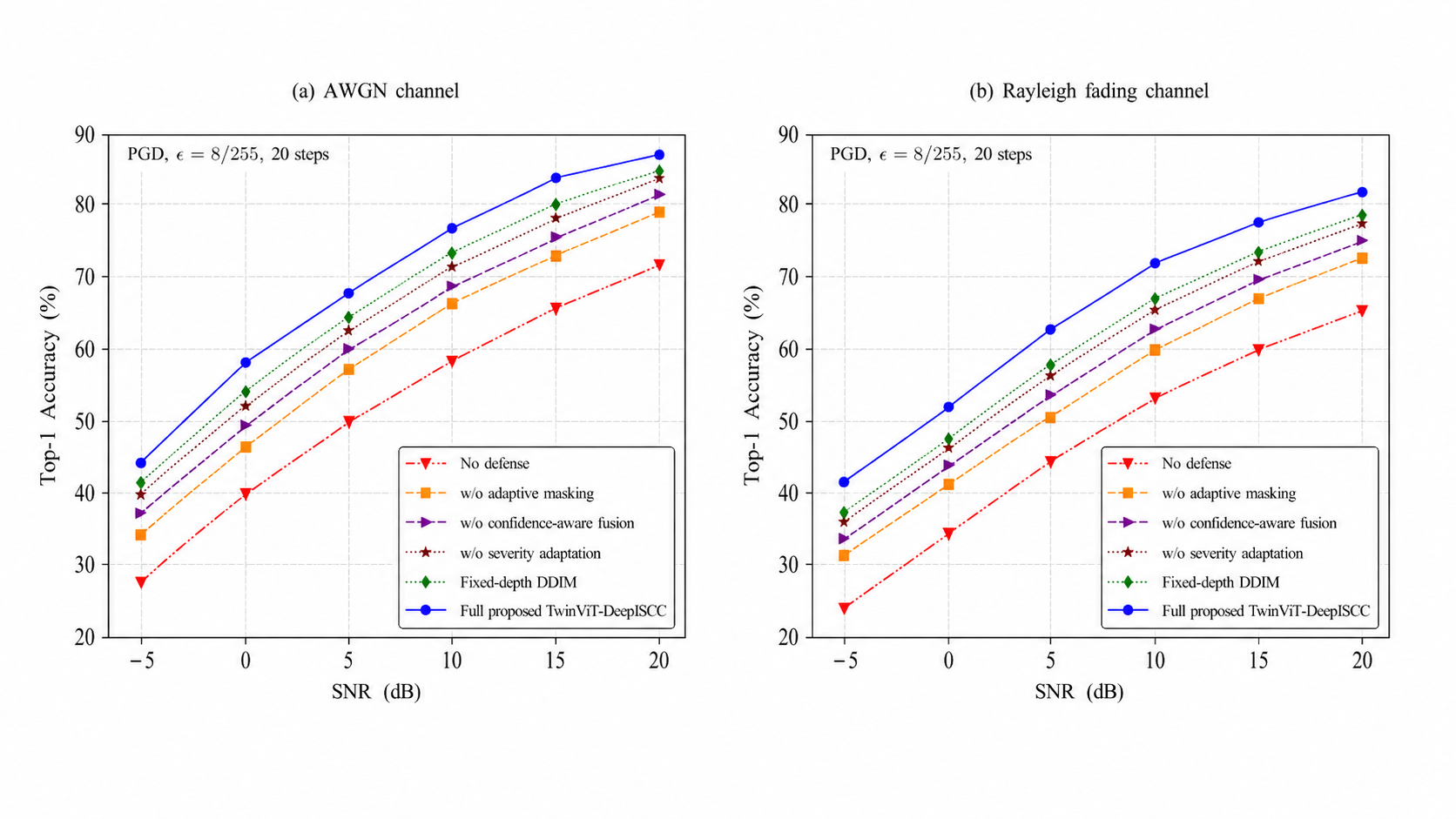}
	\caption{Module-level ablation under a 20-step PGD attack with
		$\epsilon_x=8/255$:
		(a) AWGN and
		(b) block-flat Rayleigh fading. Each incomplete variant removes or
		replaces one component while preserving the same aggregate
		channel-use budget and evaluation settings.}
	\label{fig:fig6_ablation}
\end{figure*}

Fig.~\ref{fig:fig6_ablation} isolates the contributions of adaptive latent
masking, confidence-aware fusion, and severity-adaptive diffusion
purification under a 20-step PGD attack with $\epsilon_x=8/255$.

Removing adaptive masking exposes adversarially sensitive latent
components before transmission. Replacing confidence-aware fusion with
equal-weight fusion prevents the receiver from reducing the contribution
of a branch inferred to be less reliable. Similarly, disabling severity
adaptation or using a fixed DDIM depth prevents the purification strength
from being matched to the observed corruption and channel quality.

The complete framework achieves the highest Top-1 accuracy under both
AWGN and block-flat Rayleigh fading. No individual module accounts for the entire
performance gain. Rather, the ablation trends support the complementary
operation of transmitter-side latent protection, receiver-side
reliability-aware fusion, blind severity estimation, and adaptive
diffusion purification.
\subsection{Architectural Ablation and Design Choices}
\label{subsec:architectural_ablation}

\begin{figure*}[!t]
	\centering
	\includegraphics[width=0.92\textwidth]{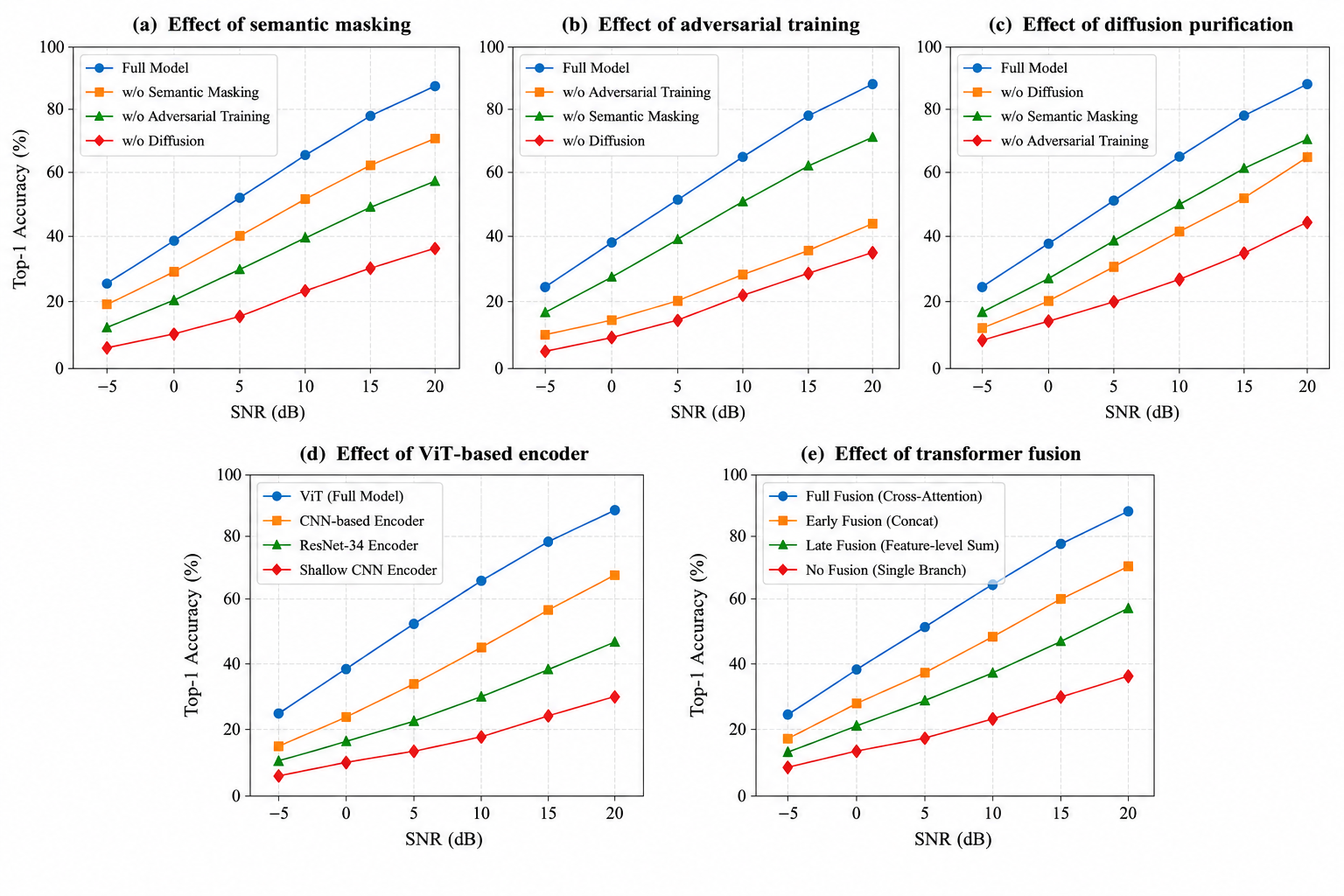}
	\caption{Architectural ablation under a 20-step PGD attack with
		$\epsilon_x=8/255$ over AWGN:
		(a) adaptive masking,
		(b) adversarial augmentation,
		(c) diffusion purification,
		(d) semantic encoder architecture, and
		(e) latent-fusion strategy.}
	\label{fig:fig7_architecture}
\end{figure*}

Fig.~\ref{fig:fig7_architecture} investigates the effects of the principal
defense mechanisms, semantic encoder backbone, and latent-fusion strategy
under identical communication resources and attack conditions.

The defense-module comparisons show that adaptive masking, adversarial
augmentation, and diffusion purification provide complementary gains.
Removing any one of these mechanisms reduces semantic accuracy, supporting
the coordinated preventive--corrective design rather than attributing the
improvement to a single isolated component.
Under the evaluated architecture settings, the ViT-based semantic encoder
outperforms the considered convolutional neural network (CNN) alternatives. This result suggests that
global feature interaction is beneficial when local image structures are
simultaneously affected by adversarial perturbation and stochastic
channel distortion. The result should, however, be interpreted within the
architectures and training configurations evaluated in this study rather
than as a universal superiority claim for ViT models.
Confidence-aware fusion also outperforms early concatenation, unweighted
late fusion, and single-branch transmission. This observation supports the
benefit of retaining two complementary branch observations and adapting
their contributions according to receiver-side reliability estimates.

\subsection{Overall Discussion}
\label{subsec:overall_discussion}

The numerical results show that TwinViT-DeepJSCC improves both reconstruction fidelity and task-level semantic preservation under diverse source- and channel-domain attacks, across the evaluated attack budgets, SNR values, and channel models. The ablation studies further indicate that the robustness gain results from the complementary operation of twin representation learning and adaptive masking, confidence-aware fusion, and severity-conditioned diffusion purification rather than from any single component. The adaptive diffusion scheduler also balances robustness and inference cost by assigning shallower reverse trajectories to higher-SNR or lower-severity observations and stronger purification to more corrupted conditions. The current evaluation is simulation-based and assumes accurate legitimate-channel estimates; hardware impairments, imperfect channel-state information, more complex fading and interference scenarios, and adaptive attacks targeting individual defense modules remain important directions for future work.

\FloatBarrier
\section{Conclusion}
\label{sec:conclusion}

This paper developed TwinViT-DeepJSCC, a preventive--corrective framework for robust semantic image communication under source-domain perturbations, channel-domain adversarial interference, AWGN, and block-flat Rayleigh fading. Under a fixed total channel-use budget, two ViT-based branches learn aligned yet redundancy-reduced semantic representations, and sample-adaptive masking suppresses vulnerable components before transmission. At the receiver, confidence-aware fusion, calibrated blind severity estimation, and conditional DDIM purification mitigate residual corruption without requiring the attack type or budget at inference.
CIFAR-100 simulations over AWGN and block-flat Rayleigh fading demonstrate improved reconstruction quality and Top-1 semantic accuracy relative to undefended, adversarially trained, and diffusion-only baselines across FGSM, PGD, NES, CW, random jamming, and channel-aware adversarial waveforms. The ablation studies confirm that the improvement arises from the complementary operation of the proposed modules.
Future work will consider imperfect channel-state information, hardware impairments, time- and frequency-selective channels, multi-user semantic communication, adaptive adversaries, and lightweight diffusion architectures suitable for real-time software-defined-radio validation.

\balance

\end{document}